\documentclass[letterpaper]{article} 
\usepackage{aaai25}  
\usepackage{times}  
\usepackage{helvet}  
\usepackage{courier}  
\usepackage[hyphens]{url}  
\usepackage{graphicx} 
\usepackage{natbib}  
\usepackage{caption} 
\usepackage{algorithm}
\usepackage{algorithmic}

\usepackage{multirow}
\usepackage{soul}

\usepackage{amsmath}
\usepackage{amssymb}

\usepackage{newfloat}
\usepackage{listings}
\DeclareCaptionStyle{ruled}{labelfont=normalfont,labelsep=colon,strut=off} 
\floatstyle{ruled}
\newfloat{listing}{tb}{lst}{}
\floatname{listing}{Listing}
\nocopyright 

\title{Leveraging Large Vision-Language Model as User Intent-aware Encoder for Composed Image Retrieval}
\author {
    Zelong Sun\textsuperscript{\rm 1},
    Dong Jing\textsuperscript{\rm 1},
    Guoxing Yang\textsuperscript{\rm 1, \rm 2},
    Nanyi Fei\textsuperscript{\rm 2},
    Zhiwu Lu\textsuperscript{\rm 1,}\thanks{Corresponding Author}
}
\affiliations {
    \textsuperscript{\rm 1}Gaoling School of Artificial Intelligence, Renmin University of China\\
    \textsuperscript{\rm 2}MetaBrain AGI Lab, Shanghai, China\\
    zelongsun@ruc.edu.com, luzhiwu@ruc.edu.com
}

\begin{document}

\maketitle

\begin{abstract}
Composed Image Retrieval (CIR) aims to retrieve target images from candidate set using a hybrid-modality query consisting of a reference image and a relative caption that describes the user intent. 
Recent studies attempt to utilize Vision-Language Pre-training Models (VLPMs) with various fusion strategies for addressing the task.
However, these methods typically fail to simultaneously meet two key requirements of CIR: comprehensively extracting visual information and faithfully following the user intent.
In this work, we propose CIR-LVLM, a novel framework that leverages the large vision-language model (LVLM) as the powerful user intent-aware encoder to better meet these requirements.
Our motivation is to explore the advanced reasoning and instruction-following capabilities of LVLM for accurately understanding and responding the user intent.
Furthermore, we design a novel hybrid intent instruction module to provide explicit intent guidance at two levels: 
(1) The task prompt clarifies the task requirement and assists the model in discerning user intent at the task level. 
(2) The instance-specific soft prompt, which is adaptively selected from the learnable prompt pool, enables the model to better comprehend the user intent at the instance level compared to a universal prompt for all instances. 
CIR-LVLM achieves state-of-the-art performance across three prominent benchmarks with acceptable inference efficiency. We believe this study provides fundamental insights into CIR-related fields.
\end{abstract}

\section{Introduction}

Compared with conventional image retrieval~\cite{i2i,i2i2, t2i, t2i2}, which involves only a single-modality query, Composed Image Retrieval (CIR) faces greater challenges as its query comprises both visual (i.e., a reference image) and textual (i.e., a relative caption) modalities. Specifically, CIR aims to retrieve the target image according to the user intent described in the relative caption and the visual information contained in the reference image. 
To achieve the goal, two fundamental challenges are presented: (1) How to comprehensively extract visual information contained in the reference image. 
(2) How to accurately capture and understand the user intent embedded in the composed query.

\begin{figure*}[t]
\centering
\vspace{-0.1in}
  \includegraphics[width=1\textwidth]{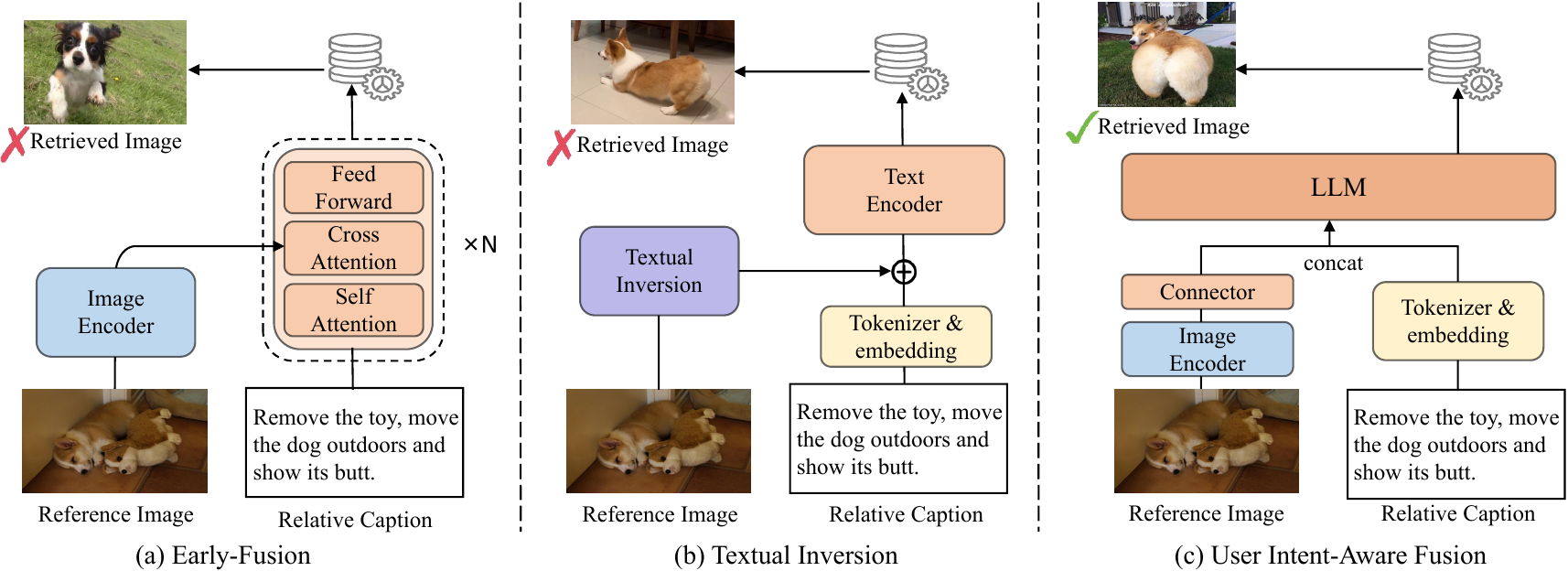}
  \caption{
  Workflows of existing CIR methods and our proposed CIR-LVLM: (a) Early-fusion, (b) Textual inversion, and (c) Our proposed CIR-LVLM.
  It can be seen that the first two fusion strategies fail to discern the user intent conveyed by the relative caption: (a) fails to retain \textit{the species of Corgi}, and (b) fails to \textit{move the dog outdoors}. Our fusion strategy leverages the superior user intent-aware capability of LVLM and successfully recalls the target image.
  }
  \label{fig:intro-way_of_fusion}
  \vspace{-0.1in}
\end{figure*}  

Most of the existing methods~\cite{clip4cir, YidaZhao2022ProgressiveLF, dataroaming, re-ranking} attempt to utilize Vision-Language Pre-training Models (VLPMs)~\cite{clip, li2022blip} with various fusion strategies for addressing these challenges.
Particularly, some methods~\cite{dataroaming, re-ranking} propose to integrate image embedding with text embedding by an unimodal encoder with intermediate cross-attention layers, known as the ``early-fusion''. However, due to the limitation of cross attention mechanism, these methods fail to retain the desired visual information, which is contained in the reference image but not mentioned in the relative caption. For example, as the visualization instance shown in Fig.\ref{fig:intro-way_of_fusion} (a), such methods tend to miss the ``species of Corgi''.
More recent approaches~\cite{saito2023pic2word, sprc} introduce a textual inversion module to generate a textual prompt from the reference image and concatenate it with the relative caption. These approaches allow a more flexible interaction for hybrid-modality queries, enabling the model to perceive more comprehensive information from the reference image. However, the limitations of the text encoder may still result in missed user intent, especially when complex relative captions are provided. As shown in Fig.\ref{fig:intro-way_of_fusion} (b), these approaches tend to ignore the user intent to ``move the dog outdoors".

Recently, several generative retrieval methods~\cite{karthik2023vision, ldre, GRB} have emerged to leverage the advanced reasoning capability of large language models (LLMs) to better perceive the user intent by inferring target image caption with LLMs from the composed query. However, the user intent captured in the inferred caption often remains at a coarse granularity level, resulting in suboptimal retrieval accuracy. Notably, these methods encounter substantial efficiency challenges due to their reliance on multi-pass decoding processes.

To overcome the limitations of the current state-of-the-art methods, we propose \textbf{CIR-LVLM}, a single-pass encoding framework designed to adopt a large vision-language model (LVLM) as a user intent-aware encoder to derive both query and target embeddings. 
As shown in Fig.\ref{fig:intro-way_of_fusion} (c), we employ the Connector module  containing a set of learnable query embeddings to generate a sentence-level prompt that enables the model to perceive comprehensive information from the image.  
Instead of using a text encoder, we leverage advanced LVLMs, which exhibit superior reasoning and instruction-following abilities, to accurately discern the user intent and obtain the desired information. 

To harness CIR-LVLM's reasoning capabilities and help it grasp CIR task patterns, we design a novel hybrid intent instruction module consisting of two kinds of prompts, providing explicit intent guidance at two levels: 
\textbf{(1) Task prompt:} 
We propose a task prompt to help the model discern user intent comprehensively at the task-level for the CIR task. 
This prompt provides detailed guidance on the task requirements, enabling the LVLM to accurately obtain the desired information from both image and text within the given context. Given the differing requirements when processing query and target images, we have designed specific task prompts for each process.
\textbf{(2) Instance-specific soft prompt:} We further employ a learnable prompt pool that automatically selects appropriate soft prompts based on the input image and text for each instance to provide Instance-level guidance. 
Different from using a universal prompt for all instances~\cite{CoOp, shin2020autoprompt}, this process allows for adaptive knowledge consolidation over changing user intent in each instance by mapping similar CIR instances onto similar prompts, maintaining the user intent-aware ability across various instance-specific requirements.

In summary, the main contributions of our paper include:\\
\textbf{(1)} To the best of our knowledge, we are the first to explore adopting the LVLM as a user intent-aware encoder in the CIR task, which accurately capture user intentions while maintaining acceptable inference efficiency\\
\textbf{(2)} We propose a novel hybrid intent instruction module tailored for the CIR task, providing explicit intent guidance at both task and instance levels.\\
\textbf{(3)} Extensive experiments conducted on three benchmarks demonstrate that CIR-LVLM outperforms the state-of-the-art CIR methods. 
Furthermore, our method can be easily transferred to the latest state-of-the-art models as they evolve, achieving even higher performance. \\
\textbf{(4)} We are the first to clearly show that in multimodal retrieval tasks that require reasoning (e.g., CIR), the LVLMs have the potential to surpass the VLPMs.

\begin{figure*}[ht!]
\centering  
\includegraphics[width=1.0\linewidth]{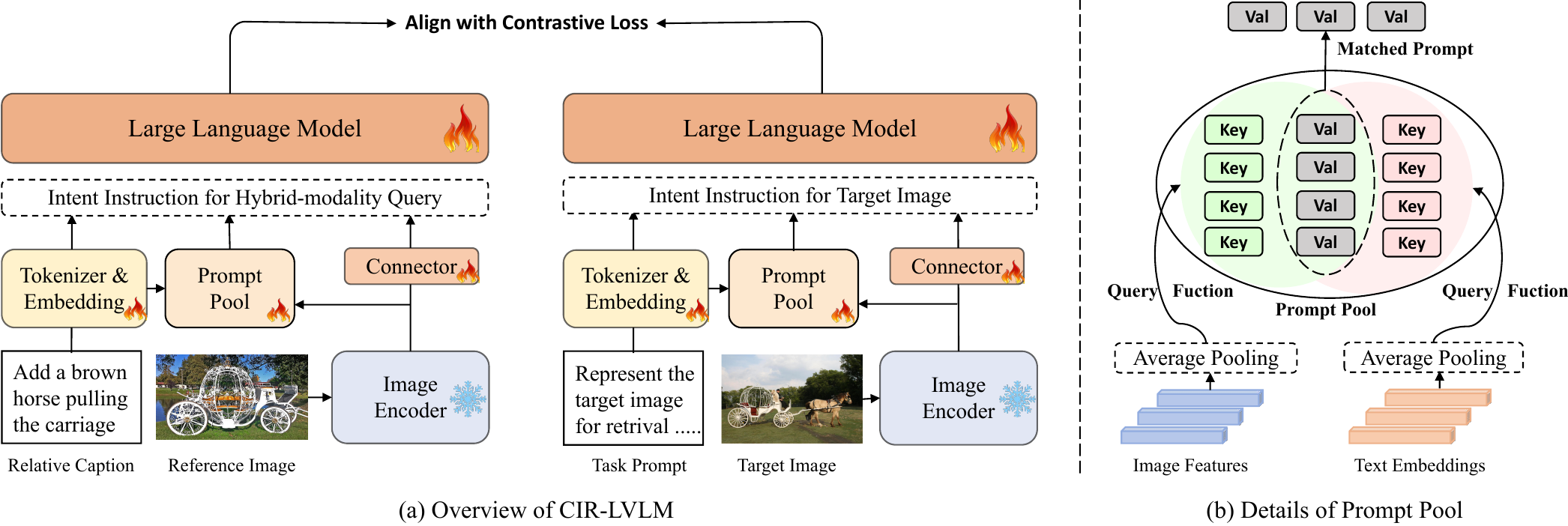} 
\vspace{-0.2in}
\caption{(a) Illustration of the architecture overview of our proposed model. All the parameters are shared between the query and target image. The intent instructions are used to form the inputs of LLM. The details of intent instructions can be found in Fig.\ref{template}. (b) Details of the prompt pool. We select prompts according to both visual features and text embeddings. }
\label{model_art}  
\vspace{-0.1in}
\end{figure*} 

\section{Related work}
\subsection{Composed image retrieval}
The prevalent contemporary CIR methods predominantly leverage the advancements in VLPMs~\cite{clip,li2022blip} as the foundational encoders and propose various strategies to adapt them to CIR task. Among them, CASE~\cite{dataroaming} and Re-ranking~\cite{re-ranking} adapt the early fusion strategy, leveraging the unimodal encoder of BLIP to fuse the information of the query. 
These methods enhance modality fusion by finer textual-visual interaction at the token-patch level.
However, since the textual-visual interaction only relies on the intermediate cross-attention layers in the unimodal encoder, these methods tend to fail to extract visual information contained in the reference image but not mentioned in the relative caption.
Another category of approaches introduces a textual inversion module to transform reference image into its pseudo-word embedding~\cite{gal2022inversion, saito2023pic2word,tang2024context} or a sentence-level prompt~\cite{sprc}, which is then concatenated with the relative caption for text-to-image retrieval. However, due to the limitation of the text encoder, the user intent is still hard to be captured. 

Recently some methods have tried to use LLM to recognize user intent for zero-shot retrieval. CIReVL~\cite{karthik2023vision} and GRB~\cite{GRB} employ a caption model to generate captions for reference images and utilize LLMs to identify user intentions and revise these captions. However, this method faces several challenges: 1. It is prone to compatibility issues between different modules. 2. The multi-pass decoding design of the caption model and LLM results in inefficiency and hallucination issues~\cite{huang2024opera, wang2023evaluation}. These limitations restrict their application in real-world scenarios. In this work, we fine-tuned the LVLM as a user intent-aware encoder to improve efficiency and reduce hallucination issues. Further details can be found in \textbf{Appendix B}.

\subsection{Large vision-language model}

With the rapid advancement of large language models~\cite{brown2020language, touvron2023llama, chatgpt}, researchers are increasingly exploring the integration of multimodal knowledge into these models. Large vision-language models (LVLMs)~\cite{ li2023blip2, liu2023llava15, dai2024instructblip, zhu2023minigpt4, bai2023qwenvl} have emerged as a prominent avenue for enhancing instruction-following capabilities through the incorporation of visual instruction tuning. 
With the unique model architecture, these models show exceptional performance across various visual tasks~\cite{agrawal2019nocaps,hudson2019gqa,mishra2019ocr}. 
In the text retrieval domain, fine-tuning LLMs as retrievers has proven to be effective, showcasing promising results~\cite{ma2023finellama, muennighoff2022sgpt, muennighoff2024representational}. However, few studies have shed light on employing LVLMs as encoders in the multimodel retrieval domain. Considering the effectiveness of LVLM in understanding and following user modification intent, in this work, we aim to explore the possibility of employing LVLM as the user intent-aware encoder in CIR task. 

\subsection{Prompt tuning}
Prompt tunings~\cite{Ptr, jin2021good, zang2022unified} is an efficient, low-cost way of adapting a pre-trained foundation model to new downstream tasks. 
In this work, We devise the task (hard) prompt to clarify the task requirement of CIR and add soft prompts to further stimulate the reasoning ability of LLMs. 
Prior related efforts include L2P~\cite{prompt_pool}, which demonstrates the potential of learnable prompts stored in a shared pool to enable continual learning without a rehearsal buffer. We also use a prompt pool to allocate soft prompts for each instance. Considering the multimodal input in CIR tasks, we select prompts using both visual and textual inputs. Such instance-specific soft prompts, compared to a universal prompt for all instances~\cite{CoOp, shin2020autoprompt}, allow the model to focus on user intent at the instance level.

\section{Method}
\label{section:method}
\subsection{Preliminary}

\textbf{Task definition.} Given a hybrid-modality query $Q=\{I_r, T\}$, where $I_r$ denotes the reference image and $T$ denotes the relative caption, and a candidate set $D=\{I_t^1, I_t^2, ..., I_t^{N_D}\}$ consisting of $N_D$ images, the goal of CIR is to identify the $k$ target images from the candidate set $D$ that are most relevant to the query $Q$, with $k \ll N_D$. 



\noindent \textbf{Challenge.} 
The core challenge of the CIR task is to modify the image content based on the relative caption while retaining as much of the reference image content as possible. Achieving such precise modifications requires the model to not only have a comprehensive understanding of the image but also possess strong reasoning abilities to accurately discern the user's intended modifications.

In this work, we aim to leverage the advanced reasoning and instruction comprehension capabilities of LVLMs to accurately capture user intent and enhance performance on CIR tasks, while maintaining acceptable inference efficiency. 
We introduce CIR-LVLM, a novel framework that fine-tunes the LVLM to function as a user intent-aware encoder, extracting representative embeddings of both hybrid-modality queries and candidate images.
Thus, CIR-LVLM need to meets three key challenges: \textbf{(1)} How to effectively extract and process visual features? \textbf{(2)} How to accurately understand the complex user requirements in CIR tasks? \textbf{(3)} How to effectively aggregate and utilize the output features of LVLM?

\subsection{Model architecture}

\textbf{Overview.}  
As shown in Fig.\ref{model_art}, we deploy the Connector containing a set of learnable query embeddings to adaptively capture the desired visual content and map it into a sentence-level prompt. This component ensures that the LLM can comprehensively perceive and understand the visual information. Then, we use LLM to identify implicit user intent from images and text, obtaining rich-content embeddings. 
However, since CIR tasks involve three inputs and are inherently complex, fully leveraging the model's ability to infer user intentions requires the model to accurately understand the relationships among these inputs. To achieve this, we propose a novel hybrid intent instruction module that includes both a task prompt and an instance-specific soft prompt, providing two levels of guidance. 
Finally, to facilitate the LLM's transition from a generative model to an encoder, we experimented with three pooling strategies to aggregate features.
In the subsequent sections, we will delve deeper into the details of these components.

\begin{figure}[t]
\centering
\includegraphics[width=0.95\linewidth]{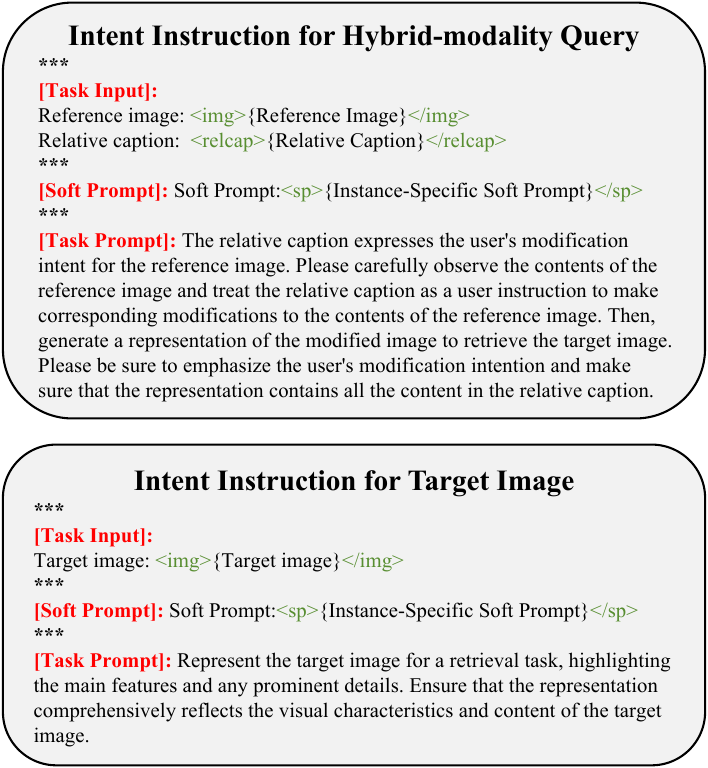}
\caption{Illustration of intent instructions for the hybrid-modality query and the target image. An intent instruction consists of three components: (1) Task Input, (2) Task Prompt, and (3) Instance-Specific Soft Prompt.}
\label{template}
\vspace{-0.1in}
\end{figure}

\noindent\textbf{Hybrid Intent instruction Module.}
CIR is a complex task that imposes different requirements to the model when processing queries and target images: deriving query embedding involves integrating information from both the reference image and relative caption, while deriving target embedding requires comprehensive extraction of visual details from the target image. 
To address these complexities and bolster the CIR-LVLM's understanding of the CIR task requirements, we design two hybrid intent instructions tailored for this context, offering explicit guidance from perspectives of task and instance.
As depicted in Fig.\ref{template}, this module emphasizes three critical elements for crafting effective prompts: Task Input, Task Prompt and Instance-Specific Soft Prompt. 
For hybrid-modality query, task input includes a reference image and a relative caption.
When processing a target image, the task input consists solely of the target image.

\textbf{Task prompt.}
The task prompt is designed to provide task-level intent guidance, clarifying the task requirements and aiding the model in comprehensively discerning user intent at the task level. Importantly, we have crafted distinct task prompts for the hybrid-modality query and the target image to cater to their respective specific needs. A detailed and well-crafted prompt can help the model better utilize its reasoning capabilities, learn higher-level concepts, and obtain more accurate representations for the input.


\begin{table*}[t!]
\fontsize{9pt}{11pt}\selectfont
\centering

\scalebox{1}{
\tabcolsep1.5pt
\begin{tabular}{l|ccccccccc|cccc}
\hline
\multirow{3}{*}{\textbf{Method}} & \multicolumn{9}{c|}{\textbf{FashionIQ}} & \multicolumn{4}{c}{\multirow{2}{*}{\textbf{Shoes}}} \\ \cline{2-10}
 & \multicolumn{2}{c|}{\textbf{Shirt}} & \multicolumn{2}{c|}{\textbf{Dress}} & \multicolumn{2}{c|}{\textbf{Tops\&Tees}} & \multicolumn{3}{c|}{\textbf{Avg.}} & \multicolumn{4}{c}{} \\ \cline{2-14} 
 & R@10 & \multicolumn{1}{c|}{R@50} & R@10 & \multicolumn{1}{c|}{R@50} & R@10 & \multicolumn{1}{c|}{R@50} & R@10 & R@50 & $R_{mean}$ & R@1 & R@10 & \multicolumn{1}{c|}{R@50} & $R_{mean}$ \\ \hline
TIRG~\cite{TIGR} & 13.10 & \multicolumn{1}{c|}{30.91} & 14.13 & \multicolumn{1}{c|}{34.61} & 14.79 & \multicolumn{1}{c|}{34.37} & 14.01 & 33.30 & 23.66 & 12.60 & 45.45 & \multicolumn{1}{c|}{69.39} & 42.48 \\
RR~\cite{Relationship} & 18.33 & \multicolumn{1}{c|}{38.63} & 15.44 & \multicolumn{1}{c|}{38.08} & 21.10 & \multicolumn{1}{c|}{44.77} & 18.29 & 40.49 & 29.39 & 12.31 & 45.10 & \multicolumn{1}{c|}{71.45} & 42.95 \\
VAL~\cite{VAL} & 22.38 & \multicolumn{1}{c|}{44.15} & 22.53 & \multicolumn{1}{c|}{44.00} & 27.53 & \multicolumn{1}{c|}{51.68} & 24.15 & 46.61 & 35.38 & 17.18 & 51.52 & \multicolumn{1}{c|}{75.83} & 48.18 \\
CoSMo~\cite{CoSMo} & 24.90 & \multicolumn{1}{c|}{49.18} & 25.64 & \multicolumn{1}{c|}{50.30} & 29.21 & \multicolumn{1}{c|}{57.46} & 26.58 & 52.31 & 39.45 & 16.72 & 48.36 & \multicolumn{1}{c|}{75.64} & 46.91 \\
CLVC-Net~\cite{CLVC-Net} & 28.75 & \multicolumn{1}{c|}{54.76} & 29.85 & \multicolumn{1}{c|}{56.47} & 33.50 & \multicolumn{1}{c|}{64.00} & 30.70 & 58.41 & 44.56 & 17.64 & 54.39 & \multicolumn{1}{c|}{79.47} & 50.50 \\
ARTEMIS~\cite{ARTEMIS} & 21.78 & \multicolumn{1}{c|}{43.64} & 27.16 & \multicolumn{1}{c|}{52.40} & 29.20 & \multicolumn{1}{c|}{54.83} & 26.05 & 50.29 & 38.17 & 18.72 & 53.11 & \multicolumn{1}{c|}{79.31} & 50.38 \\
FashionVLP~\cite{goenka2022fashionvlp} & 30.73 & \multicolumn{1}{c|}{58.02} & 30.41 & \multicolumn{1}{c|}{57.11} & 33.67 & \multicolumn{1}{c|}{64.48} & 31.60 & 59.87 & 45.735 & - & 49.08 & \multicolumn{1}{c|}{77.32} & - \\
AMC~\cite{AMC} & 30.67 & \multicolumn{1}{c|}{59.08} & 31.73 & \multicolumn{1}{c|}{59.25} & 36.21 & \multicolumn{1}{c|}{66.60} & 32.87 & 61.64 & 47.255 & 19.99 & 56.89 & \multicolumn{1}{c|}{79.27} & 52.05 \\
DCNet~\cite{DCNET} & 23.95 & \multicolumn{1}{c|}{47.30} & 28.95 & \multicolumn{1}{c|}{56.07} & 30.44 & \multicolumn{1}{c|}{58.29} & 27.78 & 53.89 & 40.84 & - & 53.82 & \multicolumn{1}{c|}{79.33} & - \\
Clip4Cir~\cite{clip4cir} & 39.99 & \multicolumn{1}{c|}{60.45} & 33.81 & \multicolumn{1}{c|}{59.40} & 41.41 & \multicolumn{1}{c|}{65.37} & 38.32 & 61.74 & 50.03 & 21.42 & 56.69 & \multicolumn{1}{c|}{81.52} & 53.21 \\
PLIR~\cite{YidaZhao2022ProgressiveLF} & 39.45 & \multicolumn{1}{c|}{61.78} & 33.60 & \multicolumn{1}{c|}{58.90} & 43.96 & \multicolumn{1}{c|}{68.33} & 39.02 & 63.00 & 51.01 & 22.88 & 58.83 & \multicolumn{1}{c|}{84.16} & 55.29 \\
CASE~\cite{dataroaming} & 48.48 & \multicolumn{1}{c|}{70.23} & 47.44 & \multicolumn{1}{c|}{69.36} & 50.18 & \multicolumn{1}{c|}{72.24} & 48.79 & 70.68 & 59.74 & - & - & \multicolumn{1}{c|}{-} & - \\
TG-CIR~\cite{TGCIR} & 52.60 & \multicolumn{1}{c|}{72.52} & 45.22 & \multicolumn{1}{c|}{69.66} & 56.14 & \multicolumn{1}{c|}{77.10} & 51.32 & 73.09 & 62.20 & {\underline{25.89}} & {\underline{63.20}} & \multicolumn{1}{c|}{{\underline{85.07}}} & {\underline{58.05}} \\
Re-ranking~\cite{re-ranking} & 50.15 & \multicolumn{1}{c|}{71.25} & 48.14 & \multicolumn{1}{c|}{71.43} & 55.23 & \multicolumn{1}{c|}{76.80} & 51.17 & 73.13 & 62.15 & - & - & \multicolumn{1}{c|}{-} & - \\
SPRC~\cite{sprc} & {\underline{55.64}} & \multicolumn{1}{c|}{{\underline{73.89}}} & {\underline{49.18}} & \multicolumn{1}{c|}{{\underline{72.43}}} & {\underline{59.35}} & \multicolumn{1}{c|}{{\underline{78.58}}} & {\underline{54.92}} & {\underline{74.97}} & {\underline{64.85}} & - & - & \multicolumn{1}{c|}{-} & - \\ 
\hline                          
\textbf{CIR-LVLM} & \textbf{58.59} & \multicolumn{1}{c|}{\textbf{75.86}} & \textbf{50.42} & \multicolumn{1}{c|}{\textbf{73.57}} & \textbf{59.61} & \multicolumn{1}{c|}{\textbf{78.99}} & \textbf{56.21} & \textbf{76.14} & \textbf{66.17} & \textbf{31.40} & \textbf{70.20} & \multicolumn{1}{c|}{\textbf{88.91}} & \textbf{63.51} \\ \hline
\end{tabular}}
\caption{Comparison with the state-of-the-art methods on the Fashion-IQ and Shoes dataset. where $R_{mean}$ indicates the average results across all the metrics. The best results are in boldface, while the second-best results are underlined.}
\vspace{-0.1in}
\label{Fashion-IQ_main}
\end{table*}

\textbf{Instance-specific soft prompt.}
The instance-specific soft prompt refines the LLM's focus toward task-specific nuances.
Previous works~\cite{CoOp, shin2020autoprompt} focus on using a universal prompt for all instances to improve the performance of pre-trained models. 
However, each instance in CIR contains subtle differences in user intent. For example, the instance shown in Fig.\ref{model_art} involves the addition of objects, whereas the instance shown in Fig.\ref{fig:intro-way_of_fusion} includes object removal, scene changes, and other complex modifications.
To address these variations and provide instance-level guidance for CIR-LVLM, we propose a shared pool of prompts that adaptively selects an instance-specific soft prompt.

Ideally, we want the model to leverage related past experiences, where similar input tend to retrieve the same group of prompts from the pool. However, since both the image and relative caption in CIR are crucial for prompt selection, we design two keys for each prompt, as shown in Fig.\ref{model_art} (b). The prompt pool is thus defined as:
\begin{equation}
    P_K = \{(k^{img}_1, k^{text}_1, P_1), \cdot \cdot \cdot,(k^{img}_M, k^{text}_M, P_M)\}
\end{equation}
where $M$ is the length of prompt pool, $P_m \in \mathbb{R}^{L_p \times d_t}$ is a learnable soft prompt with pre-defined length $L_p$ and the embedding size $d_t$, and $k^{img}_m$ and $k^{text}_m$ are learnable embeddings that represent the key for image with the shape of $\mathbb{R}^{d_i}$ and the key for text with the shape of $\mathbb{R}^{d_t}$, respectively. We denote the set of all keys by $K=\{ <k^{img}_m, k^{text}_m> \}_{i=1}^M$.

As shown in Fig.\ref{model_art} (b), our approach calculates the distance between image features and text embeddings with their corresponding keys, recalls the $top-\mathcal{K}$ closest prompts, and concatenates them in order of proximity to form an instance-specific soft prompt.
we use average pooling to aggregate the image features extracted from a frozen vision encoder as $q(I)$, where $I$ can be either a reference image $I_r$ or a target image $I_t$. Similarly, the text embedding $q(T)$ is obtained by averaging the text input embeddings of LLM. Note that $T$ represents the task prompt in Fig.\ref{template} for the target image and the relative caption for the query.
We then get a subset $K_{(I,T)}$ of $top-\mathcal{K}$ keys selected from $K$ by simply solving the objective:
\begin{equation}
K_{(I,T)} = { \underset{\{s_i\}_{i=1}^{\mathcal{K}} \in [1:M]}{{\arg\min}} \sum_{i=1}^{\mathcal{K}} \Upsilon(q(I), k_{s_i})+\Upsilon(q(T), k_{s_i}) }
\end{equation}
where $\Upsilon(\cdot)$ could be the cosine distance or another appropriate metric. This approach ensures that the chosen prompts are those most aligned with both the visual and textual inputs, allowing the model to respond more accurately and contextually to the given inputs.
To distinguish the instance-specific soft prompt,  we add two special tokens: \textit{$<sp>$} and \textit{$</sp>$}. Thus, the total soft prompt sequence is:
\begin{equation}
    S = <sp>[P_{s_1}], [P_{s_2}], ..., [P_{s_\mathcal{K}}]</sp> 
\end{equation}
Each instance can be assigned to multiple prompts, and through the prompt pool and query-key matching mechanism, different categories of knowledge can be learned. 

\noindent\textbf{Representation in CIR-LVLM.}
Previous work on the CIR task often uses a bi-directional encoder-only model, taking the representation of the prepended [CLS] token to represent the input. 
However, due to the causal attention mask in an auto-regressive decoder transformer, only the last token has attended to all tokens in a sequence. To account for this information mismatch, following~\cite{muennighoff2022sgpt}, we adopt a position-weighted mean pooling method:
\begin{equation}
\label{eq:weight_pooling}
\begin{gathered}
    V = \sum_{i=1}^{k}w_i LLM([t_1], ..., [t_k])[i] \mbox{,~where~} w_i = \frac{i}{\sum_{j=1}^{k} j}
\end{gathered}
\end{equation}
where $k$ is the sequence length, $LLM(\cdot)$ represents the LLM decoder and $w_i$ represents the positional weight which ensures that tokens appearing later have higher weights. $\{ [t_1], [t_2], ..., [t_k]$\} is input sequence that integrated with the intent instructions. 
We also experimented with last token pooling and regular mean pooling. The differences between them and the results can be found in \textbf{Appendix C}.

\noindent\textbf{Learning objective.}
The goal of training our model for CIR is to match the representation $V_Q$ of the hybrid-modality query $(I_r, T)$ with the representation $V_{I_t}$ of the target image $I_t$. At each iteration, we have a mini-batch $\{(V_Q^{(i)}, V_{I_t}^{(i)})\}_{i=1}^{N_B}$, where $(V_Q^{(i)}, V_{I_t}^{(i)})$ denotes the $i$-th pair of (hybrid-modality query, target image), and ${N_B}$ denotes the mini-batch size. Following \cite{TIGR}, we define the batch-based classification loss for model training:
\begin{equation}
L=\frac{1}{N_B}\sum_{i=1}^{N_B} -\log \frac{\exp(\lambda * Sim(V_Q^{(i)}, V_{I_t}^{(i)}) )}{\sum_{j=1}^{N_B} \exp( \lambda * Sim(V_Q^{(j)}, V_{I_t}^{(j)}) )} 
\end{equation}
where $Sim(\cdot)$ denotes the cosine similarity function, and $\lambda$ denotes a temperature parameter.

\section{Experiments}
\subsection{Datasets and evaluation metrics}
\label{section:datasets}
We make performance evaluations on three CIR benchmarks, including two fashion-domain datasets \textbf{Fashion-IQ} \cite{HuiWu2019FashionIA} and \textbf{Shoes} \cite{XiaoxiaoGuo2018DialogbasedII}, as well as an open-domain dataset \textbf{CIRR}~\cite{CIRPLANT}.

Following previous works \cite{ARTEMIS}, we adopt the Recall@K (R@K) as the evaluation metric, which refers to the fraction of queries for which the correct item is retrieved among the top K results. We also report $R_{mean}$, the mean of all R@K values, to evaluate the overall retrieval performance for Fashion-IQ and Shoes datasets. For CIRR dataset, thanks to its unique design, we can additionally report Recall$_{subset}$@K where the task is to retrieve the correct image from six curated samples, and the average score of Recall@5 and Recall$_{subset}$@1 as in \cite{CIRPLANT}.

\subsection{Implementation details}
\label{section:implementation}
We initialize our models with the Qwen-VL-Chat checkpoint and train on 8 × 80G A800 GPUs.
A challenge in fine-tuning LLMs for retrieval is the high GPU memory costs associated with contrastive learning. To address this, we employ recent memory efficiency solutions, including LoRA~\cite{hu2021lora}, flash attention~\cite{dao2023flashattention}, and gradient checkpointing to reduce GPU memory usage.

We train our model for a maximum of 10 epochs with the Adam~\cite{Adam} optimizer (with the learning rate 6e-4 for the prompt pool and 2e-4 for the rest). The temperature parameter and batch size are tailored for each dataset: 100 and 512 for Fashion-IQ, 130 and 800 for CIRR, and 70 and 512 for Shoes. 
For the parameters $L_p, \mathcal{K}, M$, we set these to 5, 8, and 45 respectively for the Fashion-IQ and Shoes datasets, and to 5, 12, and 55 for the CIRR dataset. 

\begin{table}[t!]
\center
\fontsize{9pt}{11pt}\selectfont
\scalebox{0.95}{
\tabcolsep2pt
\begin{tabular}{l|llllllll}
\hline
 & \multicolumn{4}{c}{Recall@K} & \multicolumn{3}{c}{Recall$_{subset}$@K} & \multicolumn{1}{c}{} \\ \cline{2-8}
\multirow{-2}{*}{Method} & \multicolumn{1}{c}{K=1} & \multicolumn{1}{c}{K=5} & \multicolumn{1}{c}{K=10} & \multicolumn{1}{c}{K=50} & \multicolumn{1}{c}{K=1} & \multicolumn{1}{c}{K=2} & \multicolumn{1}{c}{K=5} & \multicolumn{1}{c}{\multirow{-2}{*}{Avg.}} \\ \hline
TIRG & 14.61 & 48.37 & 64.08 & 90.03 & 22.67 & 44.97 & 65.14 & 35.52 \\
CIRPLANT & 19.55 & 52.55 & 68.39 & 92.38 & 39.20 & 63.03 & 79.49 & 45.88 \\
ARTEMIS & 16.96 & 46.10 & 61.31 & 87.73 & 39.99 & 62.20 & 75.67 & 43.05 \\
LF-CLIP & 33.59 & 65.35 & 77.35 & 95.21 & 62.39 & 81.81 & 92.02 & 72.53 \\
CLIP4CIR & 38.53 & 69.98 & 81.86 & 95.93 & 68.19 & 85.64 & 94.17 & 69.09 \\
BLIP4CIR & 40.15 & 73.08 & 83.88 & 96.27 & 72.10 & 88.27 & 95.93 & 72.59 \\
DRA & 39.93 & 72.07 & 83.83 & 96.43 & 71.04 & 87.74 & 94.72 & 71.55 \\
CASE& 48.00 & 79.11 & 87.25 & 97.57 & 75.88 & 90.58 & 96.00 & 77.50 \\
TG-CIR & 45.25 & 78.29 & 87.16 & 97.30 & 72.84 & 89.25 & 95.13 & 75.57 \\
CoVR-BLIP & 49.69 & 78.60 & 86.77 & 94.31 & 75.01 & 88.12 & 93.16 & 76.80 \\
Re-ranking & 50.55 & 81.75 & {\underline{89.78}} & 97.18 & {\underline{80.04}} & 91.90 & 96.58 & 80.90 \\
SPRC & {\underline{51.96}} & {\underline{82.12}} & 89.74 & {\underline{97.69}} & \textbf{80.65} & {\underline{92.31}} & {\underline{96.60}} & {\underline{81.38}} \\ \hline
\textbf{CIR-LVLM} & {\textbf{53.64}} & {\textbf{83.76}} & {\textbf{90.60}} & {\textbf{97.93}} & {79.12} & {\textbf{92.33}} & {\textbf{96.67}} & {\textbf{81.44}} \\ \hline
\end{tabular}}
\caption{Comparison with the state-of-the-art methods on the CIRR dataset, where Avg. indicates the average results of Recall@5 and Recall$_{subset}$@1.  The best results are in boldface, while the second-best results are underlined.}
\vspace{-0.1in}
\label{tab:cirr}
\end{table}

\subsection{Comparison with state-of-the-art methods}

Table~\ref{Fashion-IQ_main} presents the comparative results on the  \textbf{Fashion-IQ} and  \textbf{Shoes} datasets. It can be observed that: 
\textbf{(1)} Our proposed CIR-LVLM consistently achieves the highest recall across all evaluation metrics on both the Fashion-IQ and Shoes datasets. This performance emphasizes the effectiveness of fine-tuning a LVLM as a user intent-aware encoder, and the significant impact of our intent instructions comprising both the task prompt and instance-specific soft prompt.
\textbf{(2)} Compared with TG-CIR, which is the best model among the models employing the late-fusion strategy, our model leads to an improvement of  \textbf{$3.97\%$} and  \textbf{$5.46\%$} in terms of the $R_{mean}$ metric on the Fashion-IQ and Shoes dataset, respectively. CIR-LVLM's superior performance is attributed to its ability to automatically select appropriate visual features through the Connector, enabling the LLM to capture and understand the key information in images.
\textbf{(3)} Compared to the best model among the models employing the early-fusion strategy, i.e. Re-ranking, our model achieves a \textbf{$4.02\%$} improvement in $R_{mean}$ metric on the Fashion-IQ dataset. This is mainly because CIR-LVLM is more powerful in comprehensively perceiving the information of the reference image and more sensitive to user intent conveyed in the relative caption. 
\textbf{(4)} It is worth noting that, our model still outperforms SPRC across nine evaluation metrics on the Fashion-IQ dataset, even if it applies a similar textual inversion module as CIR-LVLM. This further illustrates the powerful user intent-aware abilities of LVLM in CIR task and the effectiveness of our proposed intent instructions.

\begin{table}[t]
\small
\center

\scalebox{1}{
\tabcolsep5pt
\begin{tabular}{lccc}
\hline
\multicolumn{1}{l|}{\multirow{2}{*}{Method}} & \multicolumn{3}{c}{Avg.} \\ \cline{2-4} 
\multicolumn{1}{l|}{} & R@10 & R@50 & $R_{mean}$ \\ \hline
\multicolumn{4}{c}{Encoder} \\ \hline
\multicolumn{1}{l|}{\textbf{A.1} CLIP} & 35.78 & 54.90 & 45.34 \\
\multicolumn{1}{l|}{\textbf{A.2} BLIP-2} & 43.32 & 66.39 & 54.85 \\ \hline
\multicolumn{4}{c}{Task Prompt} \\ \hline
\multicolumn{1}{l|}{\textbf{B.1} No Prompt} & 53.98 & 73.91 & 63.84 \\
\multicolumn{1}{l|}{\textbf{B.2} Brief Prompt} & 54.98 & 75.10 & 65.04 \\
\multicolumn{1}{l|}{\textbf{B.3} Detailed Prompt} & {55.82} & {75.19} & {65.50} \\ \hline
\multicolumn{4}{c}{Soft Prompt} \\ \hline
\multicolumn{1}{l|}{\textbf{C.1} Universal} & {54.80} & {75.10} & {64.95} \\ 
\multicolumn{1}{l|}{\textbf{\textbf{C.2} Instance-specific (Ours)}} & {\textbf{56.21}} & {\textbf{76.14}} & {\textbf{66.17}} \\ 
 \hline
\end{tabular}}
\caption{Ablation studies on the effects of the encoder and intent instruction. The best results are in boldface.}
\vspace{-0.1in}
\label{tab:task_template}
\end{table}

When applied to the open-domain dataset CIRR,  CIR-LVLM still demonstrates compelling results, summarized in Table~\ref{tab:cirr}. Similar conclusions can be drawn from Table~\ref{tab:cirr}, indicating that CIR-LVLM can accurately discern user intent in complex and dynamic open-world scenarios, demonstrating good generalization capabilities. However, we note that even though CIR-LVLM has made significant improvements on all other metrics,
it fails to outperform SPRC on the Recall$_{subset}$@1 metrics. This shortfall suggests that our model may struggle with distinguishing target images that are semantically or visually similar to the reference image. 

\subsection{Inference efficiency analysis}
Given that CIR models are typically employed in real-time e-commerce scenarios, maintaining acceptable inference speeds is crucial. In this section, we compare the inference speeds of CIR-LVLM with SPRC~\cite{sprc}, which uses a lightweight encoder, and CIReVL~\cite{karthik2023vision}, which also incorporates an LLM as a component. For a fair comparison, we replaced the LLM used in CIReVL with Qwen-7B instead of ChatGPT. We calculated the average time for each model to infer a single query on an A800 GPU, with the following results: SPRC took \textbf{0.035s}, CIR-LVLM took \textbf{0.08s}, and CIReVL took \textbf{1.38s}.

While CIR-LVLM's larger parameter size slightly increases inference time compared to lightweight encoders like SPRC, the difference is negligible in practical applications. In contrast, CIReVL, with a similar parameter size, significantly slows down inference due to its multi-pass decoding design. CIR-LVLM employs a single-pass encoding design, which eliminates the efficiency issues of LVLM, enabling it to be effectively applied to CIR tasks.

\section{Ablation study}
\label{section:template}
Here we conduct ablation studies on the Fashion-IQ dataset to explore the effectiveness of components in CIR-LVLM. The detailed setting and more ablation studies can be found in \textbf{Appendix C}.
\subsection{Discussion on different retrieval models.} 
To verify the effectiveness of leveraging the LLM as an encoder, in the first three rows of Table~\ref{tab:task_template}, we summarize the recalls of our method across different retrieval model configurations. It can be seen that a more powerful retrieval model leads to higher performance (refer to \textbf{A.1-2}) due to the enhanced representation capabilities. Besides, when we replace the LLM in CIR-LVLM with the text encoder of CLIP or BLIP-2, a significant decline in the $R_{mean}$ metrics could be observed. This suggests that compared to regular text encoders, the fine-tuned LLM is more advantageous in discerning user intent and extracting the desired information.

\subsection{Discussion on task prompt.} 
Here we investigate the effectiveness of the guidance from task prompt in Table~\ref{tab:task_template}. To minimize noise interference, we opted to exclude the soft prompt from all experiments. As depicted in \textbf{B.1-2} of Table~\ref{tab:task_template}, when we provide a brief prompt for the model, the model's performance improves significantly than providing no prompt. This suggests that when applying LVLM directly to the CIR task, the model will be confused about the task requirements, cause the requirements for the model are different when processing the query and the target image. Providing the model with a prompt containing the task requirements can alleviate this problem effectively. As shown in \textbf{B.3} of Table~\ref{tab:task_template}, it's evident that a more detailed task prompt leads to higher performance as it helps the model understand the CIR task more comprehensively. 
Interestingly, this impact becomes even more pronounced when LLaVA is selected as our backbone model (see \textbf{Appendix C} for details), illustrating the crucial role that prompts play when applying LVLMs to CIR tasks.

\begin{figure}[t]
\centering
\includegraphics[width=0.96\linewidth]{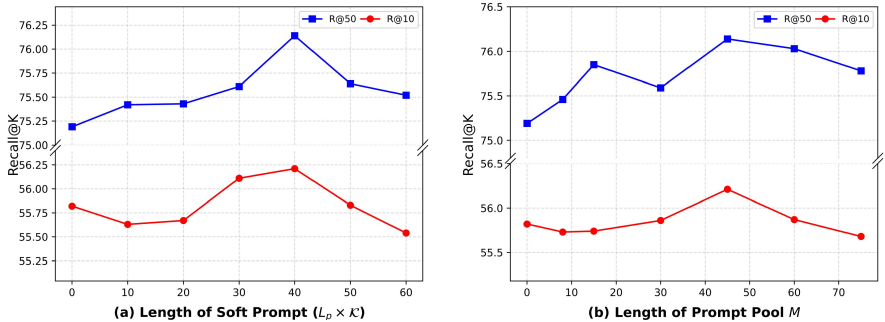}
\caption{Influence of (a) length of soft prompt and (b) length of prompt pool.}
\label{softpool}
\vspace{-0.1in}
\end{figure}

\subsection{Discussion on prompt pool.} 
We further investigate the effectiveness of the prompt pool. To achieve this, we first compare the performance of our method when applying a universal soft prompt and an instance-specific soft prompt in Table~\ref{tab:task_template} \textbf{C.1-2}. It can be observed that the instance-specific soft prompt achieved a significant improvement as it learned subtle differences between each instance and thus can provide more detailed guidance. We then discuss the effect of the length of soft prompt ($P = L_p \times \mathcal{K}$) and the length of prompt pool $M$ in Fig.\ref{softpool}. The results show that as the increase in $P$ and $M$, the recall gradually increases, indicating that the additional instance-specific soft prompt for detailed guidance is critical for CIR. Additionally, we can observe from this figure that when $P = 40$ and $M = 45$ our method obtains the highest results, after which it begins to decline due to the excessively long soft prompts or an oversized prompt pool.

\begin{figure}[t]
\centering
\includegraphics[width=0.95\linewidth]{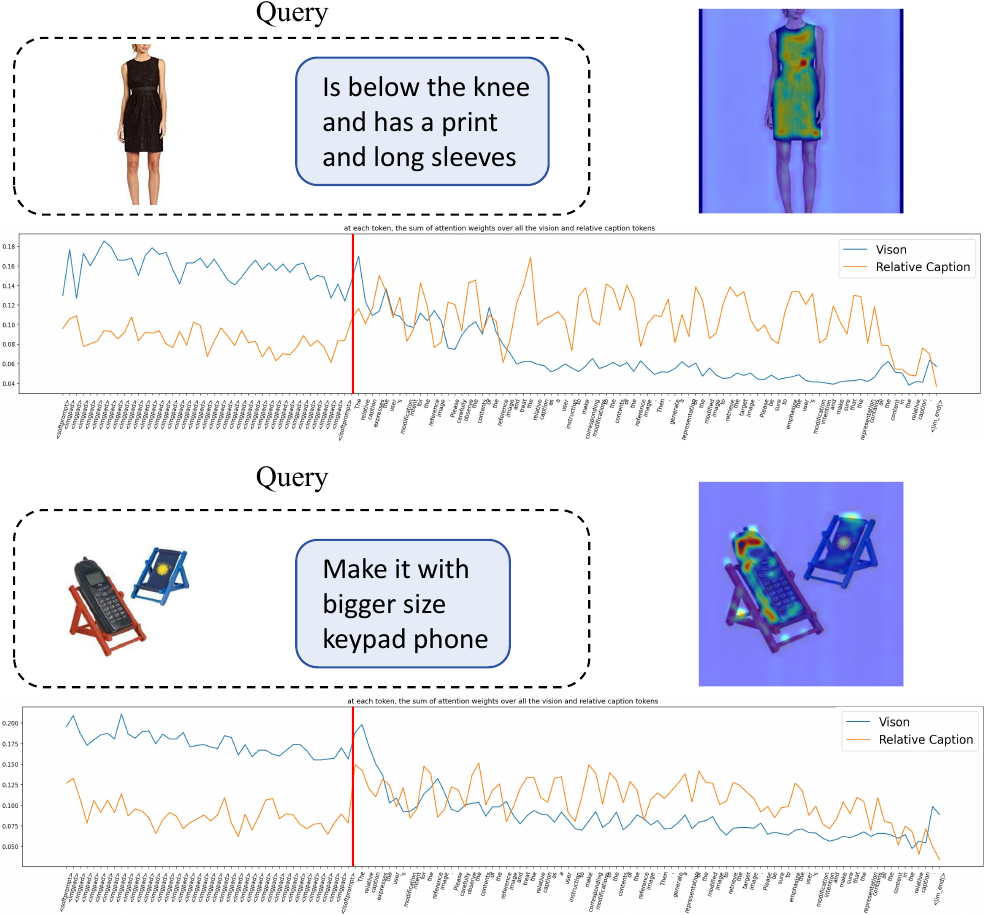}
\caption{Attention map visualization (right side of the first and third rows). The sum of the attention weights over all the visual or relative caption tokens for the soft prompt and hard prompt. (second and third rows).}
\label{attention_map}
\vspace{-0.1in}
\end{figure}

\section{Interpretability of CIR-LVLM}
In the right columns of the first and third rows of Fig.\ref{attention_map}, we visualize the attention maps of CIR-LVLM for the images. For detailed settings, please refer to the appendix. In the first example, our model successfully focuses its attention on the dress itself while largely ignoring the background. Due to the mention of ``have a print" in the relative caption, the model pays strong attention to the surface of the dress. Similarly, because the relative caption also mentions ``long sleeves", the model increases its attention to the arm area in the image. A similar conclusion can be observed in the second example, where CIR-LVLM correctly understands the user's intent and focuses on the desired image regions.

In the line charts of Fig.\ref{attention_map}, we analyze the attention weights of each token in the soft prompt and task prompt with respect to the overall visual tokens and the overall relative caption tokens. We can observe that the soft prompt places more attentions on the visual tokens, while the task prompt is more concerned about the relative caption. This conclusion is intuitive, as the task prompt primarily explains how to utilize the relative caption to accomplish the task, whereas the recall process of the soft prompt is partially driven by visual features. This finding partly explains the complementary roles these prompts play in CIR, with their combined use allowing for a more comprehensive understanding of user intent.

\section{Conclusion}
The successful application of LVLMs in various visual tasks has initiated interest in their potential to enhance retrieval tasks. In this work, we apply LVLMs to the CIR task for the first time as a user intent-aware encoder. Specifically, we utilize LVLM to provide a unified processing framework for composed querying and leverage its superior reasoning capabilities to understand and implement the user's modification intent. Furthermore, we devise novel intent instructions consisting of the task prompt and the instance-specific soft prompt to provide detailed guidance at two levels.  We are the first to clearly show that in multimodal retrieval tasks that require reasoning, the LVLMs have the potential to surpass the VLMs, offering new directions for multimodal retrieval.
\appendix
\section{Appendix}

The following items are included in our supplementary material:
\begin{itemize}
    \item Generative retrieval with LVLMs in CIR in Section~\ref{ablation:different_way}
    \item More ablation studies
    \begin{itemize}
        \item Results with different fine-tuning strategies~\ref{ablation:finetuning_startegy}
        \item Results with different backbones in Section~\ref{ablation:backbone}
        \item Results with different pooling strategies in Section~\ref{ablation:Pooling_Strategy}
        \item Detailed results with different encoders and intent instructions in Section~\ref{ablation:task_instruction}
    \end{itemize}
    \item More Visualization Experiments in Section~\ref{appdenix:attn_map}
    \item Additional qualitative analysis in Section~\ref{ablation:qualitative_analysis}
    \item Analysis of failure cases in Section~\ref{ablation:failure_case}
    \item Details of datasets in Section~\ref{ablation:datasets}
\end{itemize}

\begin{figure*}[]
\centering  
\includegraphics[width=0.75\linewidth]{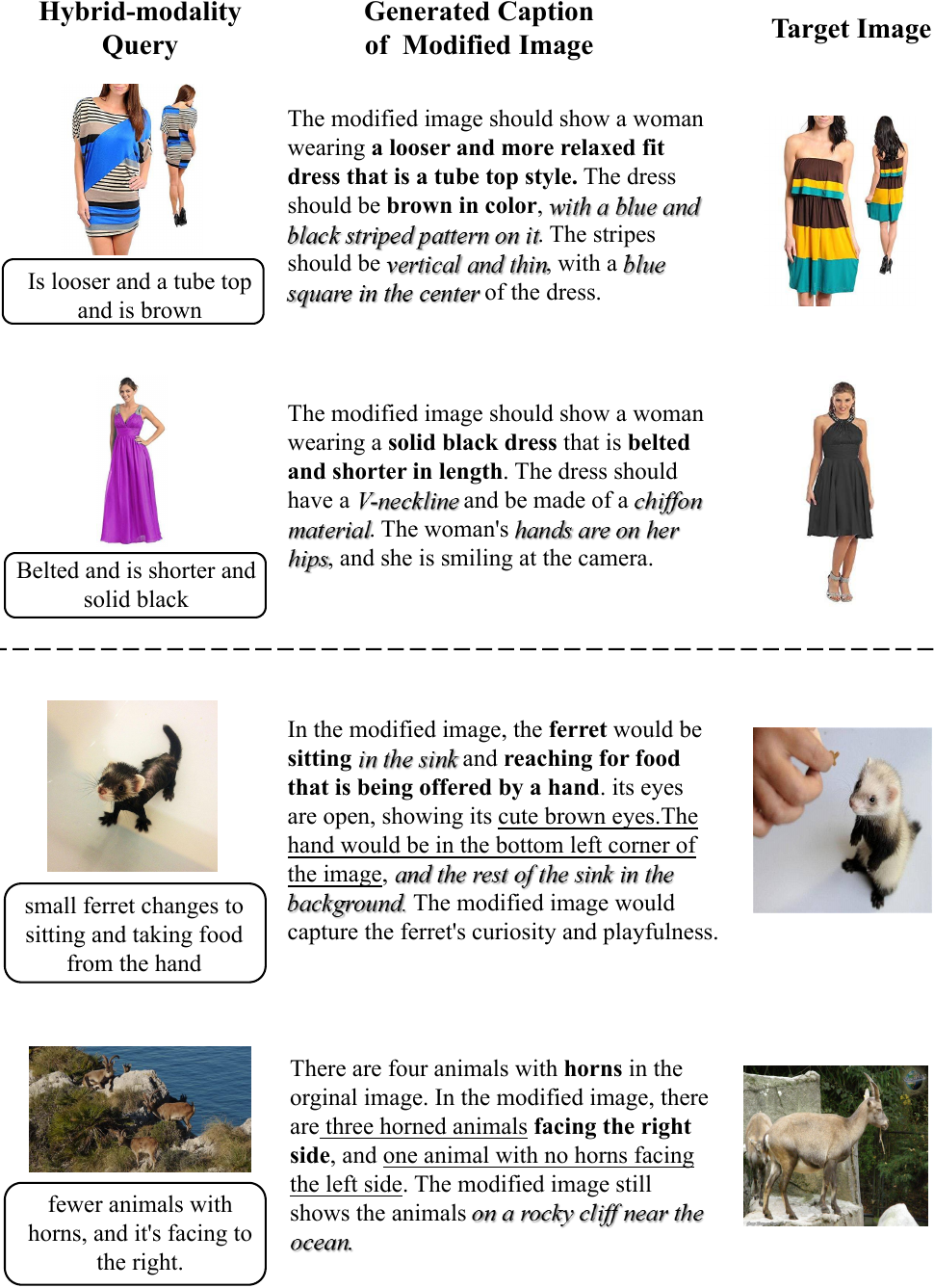} 
\caption{Generated caption of the modified version of the reference image according to the relative caption. Bolded: helpful content. Italicized: redundant information of reference image. Underlined: hallucinations.}
\label{generative_way}  
\end{figure*}

\section{Generative retrieval with LVLMs in CIR}
\label{ablation:different_way}
 Since LVLM has been trained with various generative task data, such as Caption and VQA, when we consider applying LVLM to CIR tasks, a more intuitive way is to leverage LVLM's powerful generative and reasoning capabilities to transform a mixed-modal query into a text, and then perform the querying process using a graphic retrieval model, such as CLIP. In this section, we will explain why we choose to fine-tune the LVLM to extract embeddings instead of utilizing its robust generative capability.

As shown in Fig.~\ref{generative_way}, We take the reference image and the relative caption as input to Qwen-VL-Chat and instruct it to modify the reference image based on the relative caption and generate a detailed caption of the modified image. From Fig.~\ref{generative_way} we can observe the following conclusions:

(1). \textbf{Strong visual understanding and reasoning ability}: As shown in the generated results, Qwen-VL-Chat has a strong visual understanding and is able to describe various details in the content of the image, such as ``\textbf{V-neckline}", ``\textbf{hands are on hip}" in the second example and ``\textbf{four animals}" in the fourth example. Meanwhile, as shown by the bolded text in Fig.~\ref{generative_way}, the generated answer contains all the elements in the relative caption, demonstrating the powerful reasoning and instruction-following capabilities of Qwen-VL-Chat. These capabilities are not only beneficial for integrating information from the reference image and the relative caption, but also the crucial cornerstone of the effectiveness of our work, i.e., CIR-LVLM.

(2). \textbf{Problem with redundant information of reference image}: In the CIR task, the caption of the modified image should contain both the modifications specified in the relative caption, while also preserving the remaining visual characteristics of the reference image. However, the relative captions are usually insufficient to cover all the content needed to be modified in the reference image, causing the generated captions to retain redundant information from the reference image. As shown by the italicized text with shadow in Fig.~\ref{generative_way}, these contents are real in the reference image, but create a barrier to recall the target image.

(3). \textbf{Hallucinations}: As shown by the underlined text in Fig.~\ref{generative_way}, LVLM occasionally suffers from hallucinatory problems, such as non-existent ``brown eyes", and incorrect assumptions about the position of the hands in the third example. This issue is partly due to the fact that during generation, as the sequence length grows, the self-attention increasingly focuses on the previously generated text tokens~\cite{huang2024opera, wang2023evaluation}. These mistakes are not real and pull the query away from the target image.

The hallucination issues of the information residual problem both pose a significant impediment to utilizing the generative capabilities of the LVLM in the CIR task. Besides, this approach is not efficient because of the multi-pass decoding design. 
However, fine-tuning the model into an encoder to extract the fusion features can effectively alleviate these issues, thus leading to better results. Therefore,  We chose to explore the ability of LLM to extract features in this work.

\section{More ablation studies}
\label{appendix:more_ablation}

\begin{table*}[]
\center
\scalebox{0.95}{
\tabcolsep7pt
\begin{tabular}{l|ccccccccc}
\hline
\multirow{2}{*}{Method} & \multicolumn{2}{c}{Dress} & \multicolumn{2}{c}{Shirt} & \multicolumn{2}{c}{Tops\&Tees} & \multicolumn{3}{c}{Avg.} \\ \cline{2-10} 
 & R@10 & R@50 & R@10 & R@50 & R@10 & R@50 & R@10 & R@50 & $R_{mean}$ \\ \hline
 
Zero-shot & 3.42 & 8.37 & 9.71 & 17.51 & 6.01 & 13.20 & 6.38 & 13.03 & 9.70 \\
Full Fine-tuning & 42.98 & 68.31 & 50.83 & 68.40 & 52.93 & 72.61 & 48.91 & 69.77 & 59.34 \\
LoRA w/o Connector & 47.19 & 71.04 & 57.45 & 74.92 & 56.55 & 77.40 & 53.73 & 74.46 & 64.09 \\
\textbf{Lora} & \textbf{50.42} & \textbf{73.57} & \textbf{58.59} & \textbf{75.86} & \textbf{59.64} & \textbf{78.99} & \textbf{56.21} & \textbf{76.14} & \textbf{66.17} \\ \hline
\end{tabular}}
\caption{Ablation studies on Fashion-IQ dataset with regard to different pooling strategies. The best results are in boldface.}
\label{appendix:finetun_strategy}
\end{table*}

\begin{table*}[t]
\centering
\scalebox{1}{
\tabcolsep7pt
\begin{tabular}{l|ccccccccc}
\hline
\multirow{2}{*}{Method} & \multicolumn{2}{c}{Dress} & \multicolumn{2}{c}{Shirt} & \multicolumn{2}{c}{Tops\&Tees} & \multicolumn{3}{c}{Avg.} \\ \cline{2-10} 
 & R@10 & R@50 & R@10 & R@50 & R@10 & R@50 & R@10 & R@50 & $R_{mean}$ \\ \hline
LLaVA-1.5 & 45.22 & 68.22 & 53.00 & 72.82 & 54.75 & 75.76 & 50.99 & 72.27 & 61.63 \\
LLaVA-1.6 & 44.79 & 70.43 & 54.45 & 74.33 & 54.86 & 77.04 & 51.37 & 73.93 & 62.65 \\
\textbf{Qwen-VL-Chat} & \textbf{50.42} & \textbf{73.57} & \textbf{58.59} & \textbf{75.86} & \textbf{59.64} & \textbf{78.99} & \textbf{56.21} & \textbf{76.14} & \textbf{66.17} \\ \hline
\end{tabular}}
\vspace{-0.1in}
\caption{Ablation studies on Fashion-IQ dataset with regard to different backbones. The best results are in boldface.}
\label{appendix:tab_backbones}
\end{table*}

\begin{table*}[t]
\vspace{-0.1in}
\centering
\scalebox{0.95}{
\tabcolsep7pt
\begin{tabular}{l|ccccccccc}
\hline
\multirow{2}{*}{Method} & \multicolumn{2}{c}{Dress} & \multicolumn{2}{c}{Shirt} & \multicolumn{2}{c}{Tops\&Tees} & \multicolumn{3}{c}{Avg.} \\ \cline{2-10} 
 & R@10 & R@50 & R@10 & R@50 & R@10 & R@50 & R@10 & R@50 & $R_{mean}$ \\ \hline
 Last token pooling & 50.28 & 72.48 & 57.85 & 75.57 & 59.65 & 78.37 & 55.92 & 75.47 & 65.70 \\
Regular mean pooling & 49.43 & 73.42 & 57.26 & 74.48 & 59.46 & \textbf{79.60} & 55.38 & 75.84 & 65.61 \\

\textbf{Weighted mean pooling} & \textbf{50.42} & \textbf{73.57} & \textbf{58.59} & \textbf{75.86} & \textbf{59.64} & 78.99 & \textbf{56.21} & \textbf{76.14} & \textbf{66.17} \\ \hline
\end{tabular}}
\vspace{-0.1in}
\caption{Ablation studies on Fashion-IQ dataset with regard to different pooling strategies. The best results are in boldface.}
\label{appendix:pooling_strategy}
\end{table*}

\subsection{Results with different fine-tuning strategies}
\label{ablation:finetuning_startegy}
When initiating the fine-tuning process for a LVLM, the first step involves deciding which components of the model to fine-tune. Following this, it is necessary to consider whether to update all parameters of the LLM or to employ a parameter-efficient method, such as LoRA. Table~\ref{appendix:finetun_strategy} compares the effectiveness of CIR-LVLM trained with different fine-tuning strategies.
To comprehensively explore these considerations, we delineate four distinct settings in this section:
\begin{itemize}
    \item \textbf{No fine-tuning (Zero-Shot)}: To evaluate the zero-shot capability of Qwen-VL-Chat in the CIR task, we maintain all parameters frozen during the testing phase. Here, we do not apply the prompt pool module.
    \item \textbf{Full Fine-tuning}: We freeze the vision encoder while fine-tuning the remaining components.
    \item \textbf{LoRA w/o Connector}: Following the default setting of Qwen-VL-Chat, we solely fine-tune the LLM while keeping the vision encoder and Connector frozen.
    \item \textbf{LoRA}: Given that each triplet in the CIR task comprises two images, we contend that the extraction of visual features holds particular significance. Thus, in addition to fine-tuning the LLM, we also fine-tune the Connector.
\end{itemize}

Through our experiments, we came to three conclusions: 
\textbf{(1). Zero-Shot Results}: Initially, we present the zero-shot results of CIR-LVLM (refer to the first row), revealing notably poor performance. However, following training with either full fine-tuning or LoRA, CIR-LVLM's performance exhibited remarkable improvement. This underscores the considerable potential of LLM for feature extraction tasks.
\textbf{(2). Comparison between Full Fine-Tuning and LoRA}: Upon comparing the results of full fine-tuning with LoRA (refer to the second and third row), we observe that the latter outperforms the former by approximately $6\%$ on the $R_{mean}$ metric. This suggests that full fine-tuning may be prone to overfitting the training set distribution, whereas LoRA, with significantly fewer parameters, demonstrates better generalization capabilities.
\textbf{(3). Impact of Trainable Connector in LoRA}: Notably, when keeping the Connector trainable while employing LoRA for fine-tuning (refer to the third and last row), we achieve a substantial $2.08\%$ improvement in the $R_{mean}$ metric. This underscores the critical importance of image feature extraction in the CIR task.

\subsection{Results with different backbones}
\label{ablation:backbone}

As large vision-language models (LVLMs) continue to gain prominence in the research community, we embarked on investigating the impact of different model architectures and pre-training data scales on the model's performance in the Composed Image Retrieval (CIR) task. We evaluated three distinct backbone models: \textbf{LLaVA-1.5}, \textbf{LLaVA-1.6}, and \textbf{Qwen-VL-Chat} (refer to Sections \textbf{B.1-2} and \textbf{C.1} in Table~\ref{appendix:tab_backbones} respectively). To ensure a fair comparison, all models were equipped with the LLM comprising 7 billion parameters.

\textbf{(1).} As expected, LLaVA-1.6, benefiting from higher-quality user instruction data, exhibits superior performance compared to LLaVA-1.5 in the CIR task. This underscores the notion that even when task domains differ, models pre-trained with high-quality data confer significant advantages to the CIR task. \textbf{(2).} Intriguingly, while Qwen-VL-Chat initially lags behind LLaVA-1.5 across multiple benchmarks~\cite{li2023seed, li2023evaluating, yu2023mm, liu2023mmbench}, notable improvements are observed when transitioning our backbone model from LLaVA-1.5 to Qwen-VL-Chat. We argue that this improvement primarily stems from Qwen-VL-Chat's ability to align modalities more effectively, facilitated by its cross-attention-based Connector. The utilization of 256 queries in the connector enables noise filtering in the image beforehand, enhancing alignment efficiency. Moreover, the incorporation of 256 queries helps mitigate efficiency issues associated with long image feature sequences. Consequently, we opt for Qwen-VL-Chat as our backbone model in this study.

\begin{table*}[t]
\center

\scalebox{0.95}{
\tabcolsep7pt
\begin{tabular}{lllllllccc}
\hline
\multicolumn{1}{l|}{\multirow{2}{*}{Method}} & \multicolumn{2}{c}{Dress} & \multicolumn{2}{c}{Shirt} & \multicolumn{2}{c}{Top\&Tees} & \multicolumn{3}{c}{Avg.} \\ \cline{2-10} 
\multicolumn{1}{l|}{} & \multicolumn{1}{c}{R@10} & \multicolumn{1}{c}{R@50} & \multicolumn{1}{c}{R@10} & \multicolumn{1}{c}{R@50} & \multicolumn{1}{c}{R@10} & \multicolumn{1}{c}{R@50} & R@10 & R@50 & $R_{mean}$ \\ \hline
\multicolumn{10}{c}{LLaVA-1.6} \\ \hline
\multicolumn{1}{l|}{\textbf{A.1} No Prompt} & 38.52 & 62.27 & 47.64 & 66.78 & 47.01 & 70.37 & 44.39 & 66.47 & 55.43 \\
\multicolumn{1}{l|}{\textbf{A.2} Brief Prompt} & 43.48 & 66.98 & 51.22 & 70.95 & 50.84 & 72.87 & 48.51 & 70.26 & 59.39 \\
\multicolumn{1}{l|}{\textbf{A.3} Detailed Prompt} & 44.52 & 69.16 & 52.79 & 72.08 & 53.74 & 75.52 & 50.35 & 72.25 & 61.30 \\ \hline
\multicolumn{10}{c}{Qwen-VL-chat} \\ \hline
\multicolumn{1}{l|}{\textbf{B.1} CLIP} & 32.12 & 53.29 & 38.34 & 55.86 & 36.88 & 55.54 & 35.78 & 54.90 & 45.34 \\
\multicolumn{1}{l|}{\textbf{B.2} BLIP-2} & 39.91 & 65.55 & 47.04 & 70.40 & 43.00 & 63.21 & 43.32 & 66.39 & 54.85 \\\hline
\multicolumn{1}{l|}{\textbf{C.1} No Prompt} & 47.05 & 69.98 & 57.27 & 74.37 & 57.60 & 77.63 & 53.98 & 73.91 & 63.84 \\
\multicolumn{1}{l|}{\textbf{C.2} Brief Prompt} & 50.47 & 73.12 & 57.50 & 74.92 & 57.98 & 77.97 & 55.31 & 75.34 & 65.33 \\
\multicolumn{1}{l|}{\textbf{C.3} Detailed Prompt} & 50.22 & 72.38 & 58.09 & 75.17 & 59.45 & 78.02 & 55.92 & 75.19 & 65.55 \\ \hline
\multicolumn{1}{l|}{\textbf{D.1} Universal} & 48.63 & 71.88 & 56.62 & 74.63 & 59.15 & 78.78 & 54.80 &75.10 & 64.95 \\
\multicolumn{1}{l|}{\textbf{\textbf{D.2} Instance (Ours)}} & \textbf{50.42} & \textbf{73.57} & \textbf{58.59} & \textbf{75.86} & \textbf{59.64} & \textbf{78.99} & \textbf{56.21} & \textbf{76.14} & \textbf{66.17} \\ \hline
\end{tabular}}
\caption{Detailed comparison of intent instruction in our method in terms of average recalls with regards to different task prompt and soft prompt on the Fashion-IQ dataset. The best results are in boldface.}
\label{details:task_instruction}
\end{table*}

\subsection{Results with different pooling strategies}
\label{ablation:Pooling_Strategy}
Due to the causal attention mask in an auto-regressive decoder transformer, tokens do not attend to future tokens like in an encoder transformer. Hence, only the last token has attended to all tokens in a sequence. Generally, there are three pooling strategies to acquire the input embedding: last token pooling, weighted mean pooling and regular mean pooling. In this section, we compare the performance of three pooling strategies in the CIR task.

\textbf{Last token pooling}: We append an end-of-sequence token $</s>$ to the input to form the input sequence, and use the last token as the representation of the input.  
\begin{equation}
    V = Decoder([t_1], [t_2], ..., [t_k]</s>)[-1]
\end{equation}
where $Decoder(\cdot)$ represents the Qwen-7B model, which returns the last layer token representation for each input token.

\textbf{Weighted mean pooling}: As illustrated in Equation~\ref{eq:weight_pooling}, we give later tokens a higher weight using a position-weighted mean pooling method~\cite{muennighoff2022sgpt}.

\textbf{Regular mean pooling}: Instead of giving later tokens a higher weight, we deal with these tokens equally and take their average as the representation of the input:
\begin{equation}
    V = \sum_{i=1}^{k}Decoder([t_1], [t_2], ..., [t_k])[i]
\end{equation}

The results corresponding to these settings are detailed in Table~\ref{appendix:pooling_strategy}. It can be observed that: Compared to the regular mean and last token pooling strategy, the weighted mean pooling strategy achieves significant improvements in the $R_{mean}$ metric. This suggests that the weighted mean pooling strategy is more effective cause it takes into account the importance of every token.

\begin{table*}[t]
\centering
\scalebox{0.95}{
\vspace{-0.15in}
\tabcolsep9pt
\begin{tabular}{llllllllll}
\hline
\multicolumn{1}{l|}{\multirow{2}{*}{}} & \multicolumn{2}{c}{Dress} & \multicolumn{2}{c}{Shirt} & \multicolumn{2}{c}{Top\&Tees} & \multicolumn{3}{c}{Avg.} \\ \cline{2-10} 
\multicolumn{1}{l|}{} & R@10 & R@50 & R@10 & R@50 & R@10 & R@50 & R@10 & R@50 & $R_{mean}$ \\ \hline
\multicolumn{10}{c}{Length of Soft Prompt ($L_p \times \mathcal{K}$)} \\ \hline
\multicolumn{1}{c|}{0} & 50.02 & 72.38 & 58.09 & 75.17 & 59.35 & 78.02 & 55.82 & 75.19 & 65.50 \\
\multicolumn{1}{c|}{10} & 49.82 & 72.83 & 58.68 & 74.97 & 58.38 & 78.48 & 55.63 & 75.42 & 65.48 \\
\multicolumn{1}{c|}{20} & 49.80 & 72.83 & 58.83 & 74.97 & 58.39 & 78.48 & 55.67 & 75.43 & 65.55 \\
\multicolumn{1}{c|}{30} & 50.42 & 73.52 & 58.49 & 78.63 & 59.19 & 78.47 & 56.11 & 75.61 & 65.86 \\
\multicolumn{1}{c|}{\textbf{40}} & \textbf{50.42} & \textbf{73.57} & \textbf{58.59} & \textbf{75.86} & \textbf{59.64} & \textbf{78.99} & \textbf{56.21} & \textbf{76.14} & \textbf{66.17} \\
\multicolumn{1}{c|}{50} & 50.42 & 73.23 & 58.34 & 75.03 & 58.73 & 78.67 & 55.83 & 75.64 & 65.74 \\
\multicolumn{1}{c|}{60} & 49.97 & 73.32 & 57.80 & 74.82 & 58.74 & 78.42 & 55.54 & 75.52 & 65.53 \\ \hline
\multicolumn{10}{c}{Length of Prompt Pool ($M$)} \\ \hline
\multicolumn{1}{c|}{0} & 50.02 & 72.38 & 58.09 & 75.17 & 59.35 & 78.02 & 55.82 & 75.19 & 65.50 \\
\multicolumn{1}{c|}{8} & 50.33 & 72.78 & 57.51 & 75.02 & 59.36 & 78.58 & 55.73 & 75.46 & 65.59 \\
\multicolumn{1}{c|}{15} & \textbf{50.57} & \textbf{73.77} & 57.46 & 75.17 & 59.20 & 78.63 & 55.74 & 75.85 & 65.79 \\
\multicolumn{1}{c|}{30} & 49.62 & 72.38 & 57.99 & 75.51 & \textbf{59.96} & 78.88 & 55.86 & 75.59 & 65.73 \\
\multicolumn{1}{c|}{\textbf{45}} & 50.42 & 73.57 & \textbf{58.59} & \textbf{75.86} & 59.64 & 78.99 & \textbf{56.21} & \textbf{76.14} & \textbf{66.17} \\
\multicolumn{1}{c|}{60} & 49.82 & 73.47 & 58.34 & 74.92 & 59.45 & \textbf{79.70} & 55.87 & 76.03 & 65.95 \\
\multicolumn{1}{c|}{75} & 49.92 & 73.12 & 58.19 & 75.17 & 58.94 & 79.04 & 55.68 & 75.78 & 65.68 \\ \hline
\end{tabular}}
\caption{Detailed comparison of prompt pool in our method in terms of average recalls with regards to different values of $L_p, \mathcal{K}$ and $M$ on the Fashion-IQ dataset. The best results are in boldface.}
\label{ablation:prompt_pool}
\end{table*}

\subsection{Detailed results with different encoders and intent instructions}
\label{ablation:task_instruction}

\noindent\textbf{Discussion on different retrieval models.} 
The fine-tuned LLM performs a crucial function in discerning user intent and feature extraction process. To verify the effectiveness of the fine-tuned LLM, we designed the following derivatives of CIR-LVLM:
\begin{itemize}
\item \textbf{CLIP}: We replaced the LLM with the text encoder of CLIP, while keeping the rest unchanged. Additionally, to ensure that the dimension of the sentence-level prompts matches the input dimension of CLIP's text encoder, we added an extra MLP layer after the connector module. Following CLIP's configuration, we used the [CLS] token to represent the input sequence, and obtained the final representation through CLIP's projection layer.

    \item \textbf{BLIP-2}: 
    Similar to using CLIP's text encoder, we replaced the LLM with BLIP-2's text encoder and added an extra MLP layer to complete the dimension transformation. Finally, we used the [CLS] token to represent the input sequence and obtained the final representation through the projection layer.

\end{itemize}

It can be seen that when we replace the LLM in CIR-LVLM with the text encoder of CLIP and BLIP-2, a significant decline in the $R_{mean}$ metrics could be observed. This suggests that compared to regular text ecoders, LLM is more advantageous in discerning user intent and extracting the desired information.

\noindent\textbf{Discussion on task prompt.}
To verify the importance of task prompt and prompt pool in our model, we compared CIR-LVLM with its following derivatives:
\begin{itemize}
    \item \textbf{No Prompt}: We simply concatenate the reference image's features and the relative caption's features together and feed them into the LLM. 
    \item \textbf{Brief Prompt}: We use brief hard prompt instructions, e.g., ``Describe the image as it would appear after the requested modifications have been made, focusing on the main features and any notable changes." for the query and ``Describe the image in detail, highlighting the main features and any prominent details." for the target image. 
    \item \textbf{Detailed Prompt:} As same as our full model but without the soft prompt.
\end{itemize}

As shown in Tabel~\ref{details:task_instruction}, the models that used detailed prompts achieve the best results with either Qwen-VL-chat or LLAVA-1.6 as our model backbone. It suggests that providing clear intent instructions to LVLMs proves to be beneficial.

\noindent\textbf{Discussion on prompt pool.}
We further investigate the effect of the prompt pool. To achieve this, we compared CIR-LVLM with its following derivatives:

\begin{itemize}
    \item \textbf{Universal}: Instead of the prompt pool module, we used fixed prompts with a fixed length of 40 for both the query and target image.
    \item \textbf{Instance (Ours)}: We used a prompt pool and selected soft prompts with 40 lengths from it by image features and text embedding jointly. 
    \item \textbf{Length of Soft Prompt:} To explore the effect of the length of the soft prompt ($L_p \times \mathcal{K}$), we fix $L_p$ and $M$ to $5$ and $45$ and control the length of the soft prompt by varying the value of $\mathcal{K}$.
    \item \textbf{Length of Prompt Pool: } The length of the prompt pool ($M$) has an important effect on the model. Therefore, we fix the values of $L_p$ and $\mathcal{K}$ to be $5$ and $8$ and vary the value of $M$.
\end{itemize}

As shown in Tabel~\ref{details:task_instruction} D.1-2, The fixed prompts caused a decrease in model performance instead of improving the model. This suggests that the fixed prompts do not learn the potential categorical divisions of the data, and therefore fail to provide effective detailed guidance to the model.  As shown in Table~\ref{ablation:prompt_pool}, As the length of the soft prompts and the length of the prompt pool increase, the model performance improves and achieves the best results at 40 and 45, respectively, after which the results begin to decline. To this end, we adopt $L_p=5, \mathcal{K}=8,$ and $M=45$ through our experiments.

\begin{figure*}[]
\centering  
\includegraphics[width=0.7\linewidth]{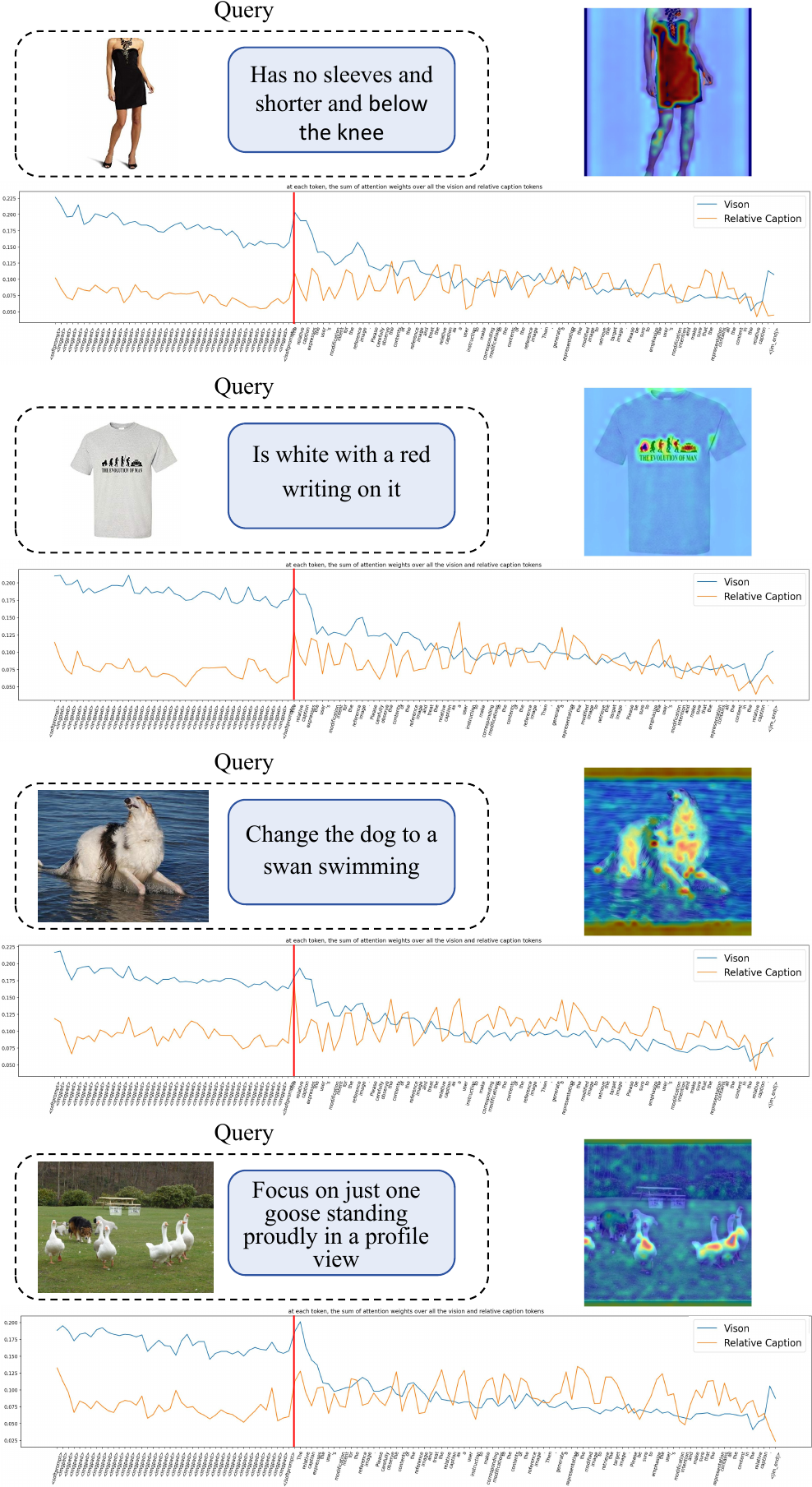} 
\caption{Attention map visualization and the sum of the attention weights over all the visual or relative caption tokens for the soft prompt and hard prompt.}
\label{appendix_attn_map}  
\vspace{-0.1in}
\end{figure*}

\section{More Visualization Experiments}
\label{appdenix:attn_map}
In Fig.~\ref{appendix_attn_map}, we present more examples of attention matrix visualizations. We have selected the attention matrices from the last layer of the vision encoder, the Connector, and the last layer of the LLM for visualization. We summed the attention matrices of tokens in the relative caption and prompt with respect to the image tokens, as CIR-LVLM employs a pooling method to aggregate the features. From the attention heatmap of the first example, we can observe that CIR-LVLM successfully focuses attention on the dress itself, and because the relative caption mentions 'no sleeves' and 'knee,' the arms and legs also receive some attention from the model. In the second example, CIR-LVLM primarily focuses on the logos and text on the clothing, which aligns with the 'red writing' mentioned in the relative caption. When there are multiple objects in the image (as in the fourth example), CIR-LVLM is able to allocate attention to each object, and because the relative caption mentions 'goose,' the dogs in the image receive significantly less attention compared to the geese. These examples clearly demonstrate CIR-LVLM's ability to accurately capture both the image information and the user intent embedded in the relative caption.

To explore the roles of the soft prompt and task prompt in CIR tasks, we also analyzed the overall attention of each prompt token towards the image and relative caption (as illustrated in the line charts in Fig.~\ref{appendix_attn_map}). Specifically, we first calculated the attention weights of each prompt token to each visual token, then summed these weights to obtain the overall attention of that prompt token to the image. The same method was applied to calculate attention towards the relative caption. We were pleasantly surprised to find that the soft prompt has a higher attention to visual tokens, while the task prompt is more focused on information within the relative caption. This conclusion is intuitive, as the task prompt primarily explains how to utilize the relative caption to accomplish the task, whereas the recall process of the soft prompt is partially driven by visual features. This finding partly explains the complementary roles these prompts play in CIR, where their combined use allows for a more comprehensive understanding of user intent.

\begin{figure*}[]
\centering  
\includegraphics[width=0.96\linewidth]{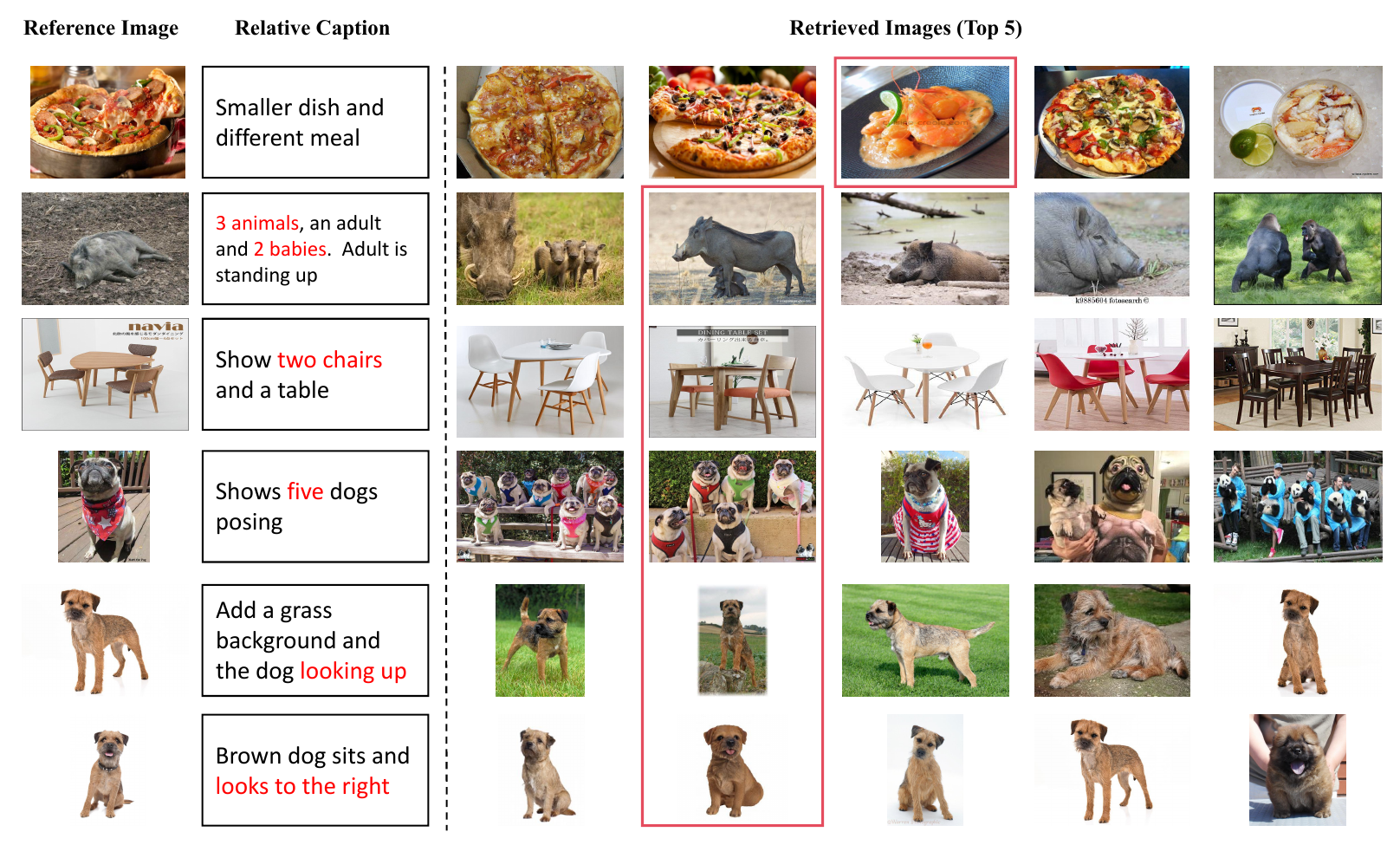} 
\caption{Failure retrieval examples obtained by our proposed model for the CIR task. Reasons for failed retrieval are highlighted in red text. The ground-truth image is highlighted with the red box.}
\label{fail_example}  
\vspace{-0.1in}
\end{figure*}  

\section{Analysis of failure cases}
\label{ablation:failure_case}
Our model shows excellent performance which is due to the strong fusion capabilities of lvlm and the guidance of the intent instructions. However, We noticed a slight drop in $Recall_s@1$ metrics in our method compared to the previous method. In this section, we analyze failure examples to guide future work on the improvement of CIR-LVLM.

As shown in Fig.~\ref{fail_example}, we have summarized three reasons that may lead to retrieval failures: \textbf{(1). Short relative captions lead to over-retention of reference image information.} We find that a large fraction of the failure cases have short relative captions. As illustrated in the first case of Fig.~\ref{fail_example}, our model ranks images with ``pizza" first, even though the relative caption contains ``different meals". A similar phenomenon can be found in the third example, where the table and chairs in the first-ranked image have the same design as in the reference image. This suggests that when the relative caption is short, the model tends to get more information about the reference image and ignore the content in the relative caption.  \textbf{(2). When the relative caption involves a change in the number of objects in the reference image, the image retrieved by CIR-LVLM usually has the correct objects, but the number does not match what the relative caption describes.} As illustrated in the second and fourth case of Fig.~\ref{fail_example}, the image with the highest recalls successfully retains the correct objects (i.e. the wild boar and the bulldog). However, the number of objects does not match the relative caption. \textbf{(2). The description of the angle change in the relative title is difficult to capture.} As shown in the last two rows in Fig.~\ref{fail_example}, the description of the angle change (i.e. ``looking up" and ``looks to the right") in the relative caption is not expressed in the top-ranked image.

\section{Additional qualitative analysis}
\label{ablation:qualitative_analysis}
\begin{figure*}[]
\centering  
\includegraphics[width=0.96\linewidth]{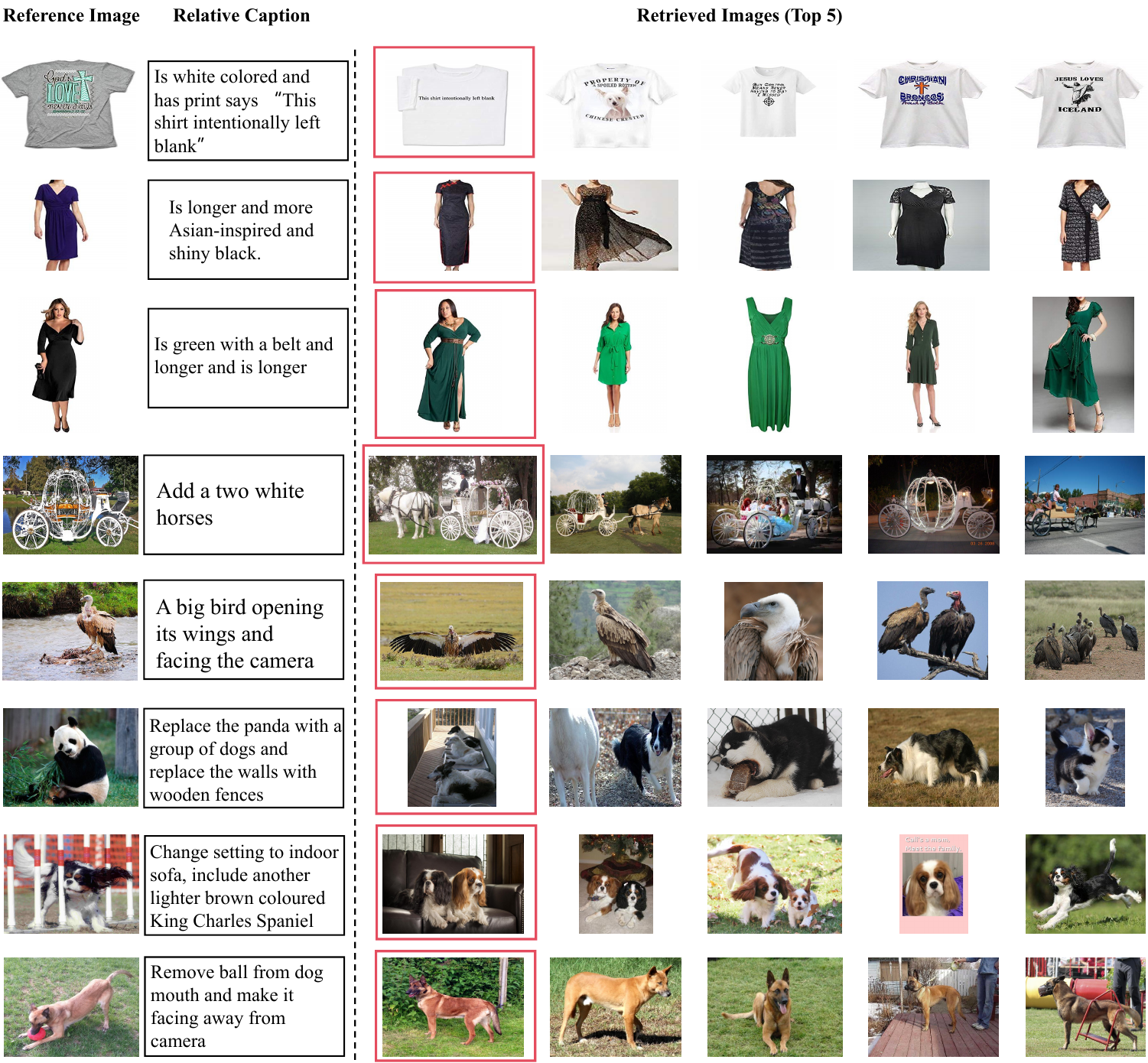} 
\vspace{-0.1in}
\caption{Successful retrieval examples obtained by our proposed model for the CIR task. The example contains operations for adding/removing/replacing objects and modifying attributes on the reference image. The ground-truth image is highlighted with the red box.}
\label{success_example}  
\end{figure*}

Fig.\ref{success_example} presents several retrieval examples from the Fashion-IQ
and CIRR datasets, where the Top-5 retrieved images are given for each hybrid-modality query, and the ground-truth image is highlighted with the red box. For a comprehensive analysis of model performance, we show examples that include operations that add, remove, replace, and modify the attributes of objects to the reference image. 
In the first row, we find that when users specify modifications to the text content on clothing, our CIR-LVLM is able to retrieve target images with the correct text content. This may be attributed to LVLM's strong OCR capabilities. In the second row, We find that our model was able to perceive changes in cultural elements, such as Asian ``Asian-inspired". In the third row, the retrieval results show that our model is able to correctly understand subtle changes such as ``a belt”.
In the fourth row, the relative caption indicates adding two horses to the reference image. Our model successfully retrieves the ground-truth image and ranks it first. Furthermore, we find that the image ranked second also has a horse and the carriage is similar in shape and color to the one in the reference image. It demonstrates that our model has learned an effective semantic embedding space for the composed image retrieval task. In the fifth row, a more detailed and descriptive modification demand is provided. Note that the relative caption does not specify the species of bird, which requires the model to fully consider both the image and relative caption in order to correctly retrieve the target image. However, the images retrieved by our model successfully retain the species of bird from the reference image and satisfy the intention from the relative caption. For the sixth row, the relative caption asks to replace the panda and the walls. It is a hard task, cause it is not easy for the fusion module to remove redundant information from the reference image. However, our model successfully eliminates the distracting information about the panda and the wall in the reference image, recalls the ground-truth image and ranks it first. The last row shows the performance of our model when facing object removal operation. Our model successfully removes the ball and retains other characteristics, e.g. background and the species of the dog in the reference image.

\vspace{-0.2in}
\section{Details of datasets}
\label{ablation:datasets}
We make performance evaluation on three benchmark, including two fashion-domain datasets \textbf{Fashion-IQ} \cite{HuiWu2019FashionIA} and \textbf{Shoes} \cite{XiaoxiaoGuo2018DialogbasedII}, as well as an open-domain dataset \textbf{CIRR}~\cite{CIRPLANT}: 
\begin{itemize}
    \item \textbf{Fashion-IQ}~\cite{HuiWu2019FashionIA} is a fashion retrieval dataset with natural language text feedback. It consists of 77,684 fashion images crawled from the \emph{Amazon.com}, which can be divided into three different categories: \emph{Dress}, \emph{Shirt}, and \emph{Tops\&tees}. Following \cite{clip4cir}, we use $18,000$ triplets for training, and $6,016$ triplets for testing. 
    
    \item \textbf{Shoes}~\cite{XiaoxiaoGuo2018DialogbasedII} is a dataset originally created for attribute discovery~\cite{TamaraLBerg2010AutomaticAD}, which is further tagged with captions for dialog-based interactive retrieval in \cite{XiaoxiaoGuo2018DialogbasedII}. It consists of 10,000 training images for training and 4,658 test images for evaluation, with 10,751 triplets annotated for IR-CQ in total.
    
    \item \textbf{CIRR}~\cite{CIRPLANT} contains 36,554 triplets, which consists of 21,552 real-life images derived from the popular natural language reasoning $NLVR^2$ dataset\cite{NLVR}. Following the standard split, we use 28,225 triplets for training, 2,297 triplets for validating, and 2,315 triplets for testing. The images of the dataset are divided into multiple subsets of six images that are semantically or visually similar. In order to have true negative images with high visual similarity, relative captions are collected so that they describe the differences between two images in the same subset. Thanks to the unique design of the CIRR dataset, we can additionally report Recall$_{subset}$@K which can be viewed as Recall while only considering images within the same subset as the pair. We also report the average score of R@5 and R$_{subset}$@1 as in \cite{CIRPLANT}.
\end{itemize}

\bibliography{aaai25}

\begin{thebibliography}{61}
\providecommand{\natexlab}[1]{#1}

\bibitem[{Agrawal et~al.(2019)Agrawal, Desai, Wang, Chen, Jain, Johnson, Batra, Parikh, Lee, and Anderson}]{agrawal2019nocaps}
Agrawal, H.; Desai, K.; Wang, Y.; Chen, X.; Jain, R.; Johnson, M.; Batra, D.; Parikh, D.; Lee, S.; and Anderson, P. 2019.
\newblock Nocaps: Novel object captioning at scale.
\newblock In \emph{Proceedings of the IEEE/CVF international conference on computer vision}, 8948--8957.

\bibitem[{Bai et~al.(2023{\natexlab{a}})Bai, Bai, Yang, Wang, Tan, Wang, Lin, Zhou, and Zhou}]{bai2023qwenvl}
Bai, J.; Bai, S.; Yang, S.; Wang, S.; Tan, S.; Wang, P.; Lin, J.; Zhou, C.; and Zhou, J. 2023{\natexlab{a}}.
\newblock Qwen-vl: A frontier large vision-language model with versatile abilities.
\newblock \emph{arXiv preprint arXiv:2308.12966}.

\bibitem[{Bai et~al.(2023{\natexlab{b}})Bai, Xu, Liu, Khan, Khan, Zuo, Goh, and Feng}]{sprc}
Bai, Y.; Xu, X.; Liu, Y.; Khan, S.; Khan, F.; Zuo, W.; Goh, R. S.~M.; and Feng, C.-M. 2023{\natexlab{b}}.
\newblock Sentence-level prompts benefit composed image retrieval.
\newblock \emph{arXiv preprint arXiv:2310.05473}.

\bibitem[{Baldrati et~al.(2022)Baldrati, Bertini, Uricchio, and Del~Bimbo}]{clip4cir}
Baldrati, A.; Bertini, M.; Uricchio, T.; and Del~Bimbo, A. 2022.
\newblock Conditioned and Composed Image Retrieval Combining and Partially Fine-Tuning CLIP-Based Features.
\newblock In \emph{IEEE/CVF Conference on Computer Vision and Pattern Recognition}, 4959--4968.

\bibitem[{Berg, Berg, and Shih(2010)}]{TamaraLBerg2010AutomaticAD}
Berg, T.~L.; Berg, A.~C.; and Shih, J. 2010.
\newblock Automatic attribute discovery and characterization from noisy web data.
\newblock In \emph{European Conference on Computer Vision}, 663--676.

\bibitem[{Brown et~al.(2020)Brown, Mann, Ryder, Subbiah, Kaplan, Dhariwal, Neelakantan, Shyam, Sastry, Askell et~al.}]{brown2020language}
Brown, T.; Mann, B.; Ryder, N.; Subbiah, M.; Kaplan, J.~D.; Dhariwal, P.; Neelakantan, A.; Shyam, P.; Sastry, G.; Askell, A.; et~al. 2020.
\newblock Language models are few-shot learners.
\newblock \emph{Advances in neural information processing systems}, 33: 1877--1901.

\bibitem[{Chen, Gong, and Bazzani(2020)}]{VAL}
Chen, Y.; Gong, S.; and Bazzani, L. 2020.
\newblock Image search with text feedback by visiolinguistic attention learning.
\newblock In \emph{IEEE/CVF Conference on Computer Vision and Pattern Recognition}, 3001--3011.

\bibitem[{Dai et~al.(2024)Dai, Li, Li, Tiong, Zhao, Wang, Li, Fung, and Hoi}]{dai2024instructblip}
Dai, W.; Li, J.; Li, D.; Tiong, A. M.~H.; Zhao, J.; Wang, W.; Li, B.; Fung, P.~N.; and Hoi, S. 2024.
\newblock Instructblip: Towards general-purpose vision-language models with instruction tuning.
\newblock \emph{Advances in Neural Information Processing Systems}, 36.

\bibitem[{Dao(2023)}]{dao2023flashattention}
Dao, T. 2023.
\newblock Flashattention-2: Faster attention with better parallelism and work partitioning.
\newblock \emph{arXiv preprint arXiv:2307.08691}.

\bibitem[{Delmas et~al.(2022)Delmas, de~Rezende, Csurka, and Larlus}]{ARTEMIS}
Delmas, G.; de~Rezende, R.~S.; Csurka, G.; and Larlus, D. 2022.
\newblock ARTEMIS: Attention-based Retrieval with Text-Explicit Matching and Implicit Similarity.
\newblock \emph{arXiv preprint arXiv:2203.08101}.

\bibitem[{Frome et~al.(2013)Frome, Corrado, Shlens, Bengio, Dean, Ranzato, and Mikolov}]{t2i}
Frome, A.; Corrado, G.~S.; Shlens, J.; Bengio, S.; Dean, J.; Ranzato, M.; and Mikolov, T. 2013.
\newblock Devise: A deep visual-semantic embedding model.
\newblock \emph{Advances in Neural Information Processing Systems}, 26: 2121--2129.

\bibitem[{Gal et~al.(2022)Gal, Alaluf, Atzmon, Patashnik, Bermano, Chechik, and Cohen-Or}]{gal2022inversion}
Gal, R.; Alaluf, Y.; Atzmon, Y.; Patashnik, O.; Bermano, A.~H.; Chechik, G.; and Cohen-Or, D. 2022.
\newblock An image is worth one word: Personalizing text-to-image generation using textual inversion.
\newblock \emph{arXiv preprint arXiv:2208.01618}.

\bibitem[{Gao et~al.(2020)Gao, Jin, Chen, Qiu, Li, Wei, Hu, and Wang}]{t2i2}
Gao, D.; Jin, L.; Chen, B.; Qiu, M.; Li, P.; Wei, Y.; Hu, Y.; and Wang, H. 2020.
\newblock Fashionbert: Text and image matching with adaptive loss for cross-modal retrieval.
\newblock In \emph{ACM SIGIR Conference on Research and Development in Information Retrieval}, 2251--2260.

\bibitem[{Goenka et~al.(2022)Goenka, Zheng, Jaiswal, Chada, Wu, Hedau, and Natarajan}]{goenka2022fashionvlp}
Goenka, S.; Zheng, Z.; Jaiswal, A.; Chada, R.; Wu, Y.; Hedau, V.; and Natarajan, P. 2022.
\newblock Fashionvlp: Vision language transformer for fashion retrieval with feedback.
\newblock In \emph{Proceedings of the IEEE/CVF Conference on Computer Vision and Pattern Recognition}, 14105--14115.

\bibitem[{Gordo et~al.(2016)Gordo, Almaz{\'a}n, Revaud, and Larlus}]{i2i}
Gordo, A.; Almaz{\'a}n, J.; Revaud, J.; and Larlus, D. 2016.
\newblock Deep image retrieval: Learning global representations for image search.
\newblock In \emph{European Conference on Computer Vision}, 241--257.

\bibitem[{Guo et~al.(2018)Guo, Wu, Cheng, Rennie, Tesauro, and Feris}]{XiaoxiaoGuo2018DialogbasedII}
Guo, X.; Wu, H.; Cheng, Y.; Rennie, S.; Tesauro, G.; and Feris, R. 2018.
\newblock Dialog-based interactive image retrieval.
\newblock \emph{Advances in Neural Information Processing Systems}, 31: 676--686.

\bibitem[{Han et~al.(2022)Han, Zhao, Ding, Liu, and Sun}]{Ptr}
Han, X.; Zhao, W.; Ding, N.; Liu, Z.; and Sun, M. 2022.
\newblock Ptr: Prompt tuning with rules for text classification.
\newblock \emph{AI Open}, 3: 182--192.

\bibitem[{Hu et~al.(2021)Hu, Shen, Wallis, Allen-Zhu, Li, Wang, Wang, and Chen}]{hu2021lora}
Hu, E.~J.; Shen, Y.; Wallis, P.; Allen-Zhu, Z.; Li, Y.; Wang, S.; Wang, L.; and Chen, W. 2021.
\newblock Lora: Low-rank adaptation of large language models.
\newblock \emph{arXiv preprint arXiv:2106.09685}.

\bibitem[{Huang et~al.(2024)Huang, Dong, Zhang, Wang, He, Wang, Lin, Zhang, and Yu}]{huang2024opera}
Huang, Q.; Dong, X.; Zhang, P.; Wang, B.; He, C.; Wang, J.; Lin, D.; Zhang, W.; and Yu, N. 2024.
\newblock Opera: Alleviating hallucination in multi-modal large language models via over-trust penalty and retrospection-allocation.
\newblock In \emph{Proceedings of the IEEE/CVF Conference on Computer Vision and Pattern Recognition}, 13418--13427.

\bibitem[{Hudson and Manning(2019)}]{hudson2019gqa}
Hudson, D.~A.; and Manning, C.~D. 2019.
\newblock Gqa: A new dataset for real-world visual reasoning and compositional question answering.
\newblock In \emph{Proceedings of the IEEE/CVF conference on computer vision and pattern recognition}, 6700--6709.

\bibitem[{Jin et~al.(2021)Jin, Cheng, Shen, Chen, and Ren}]{jin2021good}
Jin, W.; Cheng, Y.; Shen, Y.; Chen, W.; and Ren, X. 2021.
\newblock A good prompt is worth millions of parameters: Low-resource prompt-based learning for vision-language models.
\newblock \emph{arXiv preprint arXiv:2110.08484}.

\bibitem[{Karthik et~al.(2023)Karthik, Roth, Mancini, and Akata}]{karthik2023vision}
Karthik, S.; Roth, K.; Mancini, M.; and Akata, Z. 2023.
\newblock Vision-by-language for training-free compositional image retrieval.
\newblock \emph{arXiv preprint arXiv:2310.09291}.

\bibitem[{Kim et~al.(2021)Kim, Yu, Kim, and Kim}]{DCNET}
Kim, J.; Yu, Y.; Kim, H.; and Kim, G. 2021.
\newblock Dual compositional learning in interactive image retrieval.
\newblock In \emph{AAAI Conference on Artificial Intelligence}, 1771--1779.

\bibitem[{Kingma and Ba(2014)}]{Adam}
Kingma, D.~P.; and Ba, J. 2014.
\newblock Adam: A method for stochastic optimization.
\newblock \emph{arXiv preprint arXiv:1412.6980}.

\bibitem[{Lee, Kim, and Han(2021)}]{CoSMo}
Lee, S.; Kim, D.; and Han, B. 2021.
\newblock Cosmo: Content-style modulation for image retrieval with text feedback.
\newblock In \emph{IEEE/CVF Conference on Computer Vision and Pattern Recognition}, 802--812.

\bibitem[{Levy et~al.(2023)Levy, Ben-Ari, Darshan, and Lischinski}]{dataroaming}
Levy, M.; Ben-Ari, R.; Darshan, N.; and Lischinski, D. 2023.
\newblock Data Roaming and Early Fusion for Composed Image Retrieval.
\newblock \emph{arXiv preprint arXiv:2303.09429}.

\bibitem[{Li et~al.(2023{\natexlab{a}})Li, Wang, Wang, Ge, Ge, and Shan}]{li2023seed}
Li, B.; Wang, R.; Wang, G.; Ge, Y.; Ge, Y.; and Shan, Y. 2023{\natexlab{a}}.
\newblock Seed-bench: Benchmarking multimodal llms with generative comprehension.
\newblock \emph{arXiv preprint arXiv:2307.16125}.

\bibitem[{Li et~al.(2023{\natexlab{b}})Li, Li, Savarese, and Hoi}]{li2023blip2}
Li, J.; Li, D.; Savarese, S.; and Hoi, S. 2023{\natexlab{b}}.
\newblock Blip-2: Bootstrapping language-image pre-training with frozen image encoders and large language models.
\newblock In \emph{International conference on machine learning}, 19730--19742. PMLR.

\bibitem[{Li et~al.(2022)Li, Li, Xiong, and Hoi}]{li2022blip}
Li, J.; Li, D.; Xiong, C.; and Hoi, S. 2022.
\newblock Blip: Bootstrapping language-image pre-training for unified vision-language understanding and generation.
\newblock In \emph{International conference on machine learning}, 12888--12900. PMLR.

\bibitem[{Li et~al.(2023{\natexlab{c}})Li, Du, Zhou, Wang, Zhao, and Wen}]{li2023evaluating}
Li, Y.; Du, Y.; Zhou, K.; Wang, J.; Zhao, W.~X.; and Wen, J.-R. 2023{\natexlab{c}}.
\newblock Evaluating object hallucination in large vision-language models.
\newblock \emph{arXiv preprint arXiv:2305.10355}.

\bibitem[{Liu et~al.(2023{\natexlab{a}})Liu, Li, Li, and Lee}]{liu2023llava15}
Liu, H.; Li, C.; Li, Y.; and Lee, Y.~J. 2023{\natexlab{a}}.
\newblock Improved baselines with visual instruction tuning.
\newblock \emph{arXiv preprint arXiv:2310.03744}.

\bibitem[{Liu et~al.(2023{\natexlab{b}})Liu, Duan, Zhang, Li, Zhang, Zhao, Yuan, Wang, He, Liu et~al.}]{liu2023mmbench}
Liu, Y.; Duan, H.; Zhang, Y.; Li, B.; Zhang, S.; Zhao, W.; Yuan, Y.; Wang, J.; He, C.; Liu, Z.; et~al. 2023{\natexlab{b}}.
\newblock Mmbench: Is your multi-modal model an all-around player?
\newblock \emph{arXiv preprint arXiv:2307.06281}.

\bibitem[{Liu et~al.(2016)Liu, Luo, Qiu, Wang, and Tang}]{i2i2}
Liu, Z.; Luo, P.; Qiu, S.; Wang, X.; and Tang, X. 2016.
\newblock Deepfashion: Powering robust clothes recognition and retrieval with rich annotations.
\newblock In \emph{IEEE Conference on Computer Vision and Pattern Recognition}, 1096--1104.

\bibitem[{Liu et~al.(2021)Liu, Rodriguez-Opazo, Teney, and Gould}]{CIRPLANT}
Liu, Z.; Rodriguez-Opazo, C.; Teney, D.; and Gould, S. 2021.
\newblock Image retrieval on real-life images with pre-trained vision-and-language models.
\newblock In \emph{Proceedings of the IEEE/CVF International Conference on Computer Vision}, 2125--2134.

\bibitem[{Liu et~al.(2023{\natexlab{c}})Liu, Sun, Teney, and Gould}]{re-ranking}
Liu, Z.; Sun, W.; Teney, D.; and Gould, S. 2023{\natexlab{c}}.
\newblock Candidate Set Re-ranking for Composed Image Retrieval with Dual Multi-modal Encoder.
\newblock \emph{arXiv preprint arXiv:2305.16304}.

\bibitem[{Ma et~al.(2023)Ma, Wang, Yang, Wei, and Lin}]{ma2023finellama}
Ma, X.; Wang, L.; Yang, N.; Wei, F.; and Lin, J. 2023.
\newblock Fine-tuning llama for multi-stage text retrieval.
\newblock \emph{arXiv preprint arXiv:2310.08319}.

\bibitem[{Mishra et~al.(2019)Mishra, Shekhar, Singh, and Chakraborty}]{mishra2019ocr}
Mishra, A.; Shekhar, S.; Singh, A.~K.; and Chakraborty, A. 2019.
\newblock Ocr-vqa: Visual question answering by reading text in images.
\newblock In \emph{2019 international conference on document analysis and recognition (ICDAR)}, 947--952. IEEE.

\bibitem[{Muennighoff(2022)}]{muennighoff2022sgpt}
Muennighoff, N. 2022.
\newblock Sgpt: Gpt sentence embeddings for semantic search.
\newblock \emph{arXiv preprint arXiv:2202.08904}.

\bibitem[{Muennighoff et~al.(2024)Muennighoff, Su, Wang, Yang, Wei, Yu, Singh, and Kiela}]{muennighoff2024representational}
Muennighoff, N.; Su, H.; Wang, L.; Yang, N.; Wei, F.; Yu, T.; Singh, A.; and Kiela, D. 2024.
\newblock Generative representational instruction tuning.
\newblock \emph{arXiv preprint arXiv:2402.09906}.

\bibitem[{OpenAI(2023)}]{chatgpt}
OpenAI. 2023.
\newblock ChatGPT.
\newblock \url{https://chat.openai.com/}.

\bibitem[{Radford et~al.(2021)Radford, Kim, Hallacy, Ramesh, Goh, Agarwal, Sastry, Askell, Mishkin, Clark et~al.}]{clip}
Radford, A.; Kim, J.~W.; Hallacy, C.; Ramesh, A.; Goh, G.; Agarwal, S.; Sastry, G.; Askell, A.; Mishkin, P.; Clark, J.; et~al. 2021.
\newblock Learning transferable visual models from natural language supervision.
\newblock In \emph{International conference on machine learning}, 8748--8763. PMLR.

\bibitem[{Saito et~al.(2023)Saito, Sohn, Zhang, Li, Lee, Saenko, and Pfister}]{saito2023pic2word}
Saito, K.; Sohn, K.; Zhang, X.; Li, C.-L.; Lee, C.-Y.; Saenko, K.; and Pfister, T. 2023.
\newblock Pic2word: Mapping pictures to words for zero-shot composed image retrieval.
\newblock In \emph{Proceedings of the IEEE/CVF Conference on Computer Vision and Pattern Recognition}, 19305--19314.

\bibitem[{Santoro et~al.(2017)Santoro, Raposo, Barrett, Malinowski, Pascanu, Battaglia, and Lillicrap}]{Relationship}
Santoro, A.; Raposo, D.; Barrett, D.~G.; Malinowski, M.; Pascanu, R.; Battaglia, P.; and Lillicrap, T. 2017.
\newblock A simple neural network module for relational reasoning.
\newblock \emph{Advances in Neural Information Processing Systems}, 30: 4974--4983.

\bibitem[{Shin et~al.(2020)Shin, Razeghi, Logan~IV, Wallace, and Singh}]{shin2020autoprompt}
Shin, T.; Razeghi, Y.; Logan~IV, R.~L.; Wallace, E.; and Singh, S. 2020.
\newblock Autoprompt: Eliciting knowledge from language models with automatically generated prompts.
\newblock \emph{arXiv preprint arXiv:2010.15980}.

\bibitem[{Suhr et~al.(2018)Suhr, Zhou, Zhang, Zhang, Bai, and Artzi}]{NLVR}
Suhr, A.; Zhou, S.; Zhang, A.; Zhang, I.; Bai, H.; and Artzi, Y. 2018.
\newblock A corpus for reasoning about natural language grounded in photographs.
\newblock \emph{arXiv preprint arXiv:1811.00491}.

\bibitem[{Sun, Ye, and Gong(2023)}]{GRB}
Sun, S.; Ye, F.; and Gong, S. 2023.
\newblock Training-free Zero-shot Composed Image Retrieval with Local Concept Reranking.
\newblock \emph{arXiv preprint arXiv:2312.08924}.

\bibitem[{Tang et~al.(2024)Tang, Yu, Gai, Zhuang, Xiong, Hu, and Wu}]{tang2024context}
Tang, Y.; Yu, J.; Gai, K.; Zhuang, J.; Xiong, G.; Hu, Y.; and Wu, Q. 2024.
\newblock Context-I2W: Mapping Images to Context-dependent Words for Accurate Zero-Shot Composed Image Retrieval.
\newblock In \emph{Proceedings of the AAAI Conference on Artificial Intelligence}, volume~38, 5180--5188.

\bibitem[{Touvron et~al.(2023)Touvron, Lavril, Izacard, Martinet, Lachaux, Lacroix, Rozi{\`e}re, Goyal, Hambro, Azhar et~al.}]{touvron2023llama}
Touvron, H.; Lavril, T.; Izacard, G.; Martinet, X.; Lachaux, M.-A.; Lacroix, T.; Rozi{\`e}re, B.; Goyal, N.; Hambro, E.; Azhar, F.; et~al. 2023.
\newblock Llama: Open and efficient foundation language models.
\newblock \emph{arXiv preprint arXiv:2302.13971}.

\bibitem[{Vo et~al.(2019)Vo, Jiang, Sun, Murphy, Li, Fei-Fei, and Hays}]{TIGR}
Vo, N.; Jiang, L.; Sun, C.; Murphy, K.; Li, L.-J.; Fei-Fei, L.; and Hays, J. 2019.
\newblock Composing text and image for image retrieval-an empirical odyssey.
\newblock In \emph{IEEE/CVF Conference on Computer Vision and Pattern Recognition}, 6439--6448.

\bibitem[{Wang et~al.(2023)Wang, Zhou, Xu, Shi, Zhao, Xu, Ye, Yan, Zhang, Zhu et~al.}]{wang2023evaluation}
Wang, J.; Zhou, Y.; Xu, G.; Shi, P.; Zhao, C.; Xu, H.; Ye, Q.; Yan, M.; Zhang, J.; Zhu, J.; et~al. 2023.
\newblock Evaluation and Analysis of Hallucination in Large Vision-Language Models.(Aug.
\newblock \emph{arXiv preprint arxiv:2308.15126}.

\bibitem[{Wang et~al.(2022)Wang, Zhang, Lee, Zhang, Sun, Ren, Su, Perot, Dy, and Pfister}]{prompt_pool}
Wang, Z.; Zhang, Z.; Lee, C.-Y.; Zhang, H.; Sun, R.; Ren, X.; Su, G.; Perot, V.; Dy, J.; and Pfister, T. 2022.
\newblock Learning to prompt for continual learning.
\newblock In \emph{Proceedings of the IEEE/CVF Conference on Computer Vision and Pattern Recognition}, 139--149.

\bibitem[{Wen et~al.(2021)Wen, Song, Yang, Zhan, and Nie}]{CLVC-Net}
Wen, H.; Song, X.; Yang, X.; Zhan, Y.; and Nie, L. 2021.
\newblock Comprehensive linguistic-visual composition network for image retrieval.
\newblock In \emph{ACM SIGIR Conference on Research and Development in Information Retrieval}, 1369--1378.

\bibitem[{Wen et~al.(2023)Wen, Zhang, Song, Wei, and Nie}]{TGCIR}
Wen, H.; Zhang, X.; Song, X.; Wei, Y.; and Nie, L. 2023.
\newblock Target-guided composed image retrieval.
\newblock In \emph{Proceedings of the 31st ACM International Conference on Multimedia}, 915--923.

\bibitem[{Wu et~al.(2021)Wu, Gao, Guo, Al-Halah, Rennie, Grauman, and Feris}]{HuiWu2019FashionIA}
Wu, H.; Gao, Y.; Guo, X.; Al-Halah, Z.; Rennie, S.; Grauman, K.; and Feris, R. 2021.
\newblock Fashion iq: A new dataset towards retrieving images by natural language feedback.
\newblock In \emph{IEEE/CVF Conference on Computer Vision and Pattern Recognition}, 11307--11317.

\bibitem[{Yang et~al.(2024)Yang, Xue, Qian, Dong, and Xu}]{ldre}
Yang, Z.; Xue, D.; Qian, S.; Dong, W.; and Xu, C. 2024.
\newblock LDRE: LLM-based Divergent Reasoning and Ensemble for Zero-Shot Composed Image Retrieval.
\newblock In \emph{SIGIR}, 80--90.

\bibitem[{Yu et~al.(2023)Yu, Yang, Li, Wang, Lin, Liu, Wang, and Wang}]{yu2023mm}
Yu, W.; Yang, Z.; Li, L.; Wang, J.; Lin, K.; Liu, Z.; Wang, X.; and Wang, L. 2023.
\newblock Mm-vet: Evaluating large multimodal models for integrated capabilities.
\newblock \emph{arXiv preprint arXiv:2308.02490}.

\bibitem[{Zang et~al.(2022)Zang, Li, Zhou, Huang, and Loy}]{zang2022unified}
Zang, Y.; Li, W.; Zhou, K.; Huang, C.; and Loy, C.~C. 2022.
\newblock Unified vision and language prompt learning.
\newblock \emph{arXiv preprint arXiv:2210.07225}.

\bibitem[{Zhao, Song, and Jin(2022)}]{YidaZhao2022ProgressiveLF}
Zhao, Y.; Song, Y.; and Jin, Q. 2022.
\newblock Progressive Learning for Image Retrieval with Hybrid-Modality Queries.
\newblock \emph{arXiv preprint arXiv:2204.11212}.

\bibitem[{Zhou et~al.(2022)Zhou, Yang, Loy, and Liu}]{CoOp}
Zhou, K.; Yang, J.; Loy, C.~C.; and Liu, Z. 2022.
\newblock Learning to prompt for vision-language models.
\newblock \emph{International Journal of Computer Vision}, 130(9): 2337--2348.

\bibitem[{Zhu et~al.(2023{\natexlab{a}})Zhu, Chen, Shen, Li, and Elhoseiny}]{zhu2023minigpt4}
Zhu, D.; Chen, J.; Shen, X.; Li, X.; and Elhoseiny, M. 2023{\natexlab{a}}.
\newblock Minigpt-4: Enhancing vision-language understanding with advanced large language models.
\newblock \emph{arXiv preprint arXiv:2304.10592}.

\bibitem[{Zhu et~al.(2023{\natexlab{b}})Zhu, Wei, Zhao, Zhang, and Huang}]{AMC}
Zhu, H.; Wei, Y.; Zhao, Y.; Zhang, C.; and Huang, S. 2023{\natexlab{b}}.
\newblock Amc: Adaptive multi-expert collaborative network for text-guided image retrieval.
\newblock \emph{ACM Transactions on Multimedia Computing, Communications and Applications}, 19(6): 1--22.

\end{thebibliography}

\end{document}